\documentclass[epj]{svjour}
\usepackage{graphics}
\usepackage{amsfonts}
\usepackage{epsfig}
\usepackage{graphicx,amsmath,amssymb}
\newcommand{\bb}{\begin{equation}}
\newcommand{\en}{\end{equation}}

\def\vr{{\bf r}}
\def\vf{{\bf f}}
\def\vv{{\bf v}}
\def\rp{{\bf r_\perp}}

\def\kp{{\bf k}_\perp}
\def\kps{k_\perp}
\def\vp{{\bf v_\perp}}
\def\fp{{\bf f_\perp}}
\def\vn{{\bf \hat{n}}}
\def\vz{{\bf \hat{z}}}
\def\Ep{{\bf E_\perp}}

\def\kph{{\bf \hat{k}}_\perp}
\def\vt{{\bf \hat{t}}}

\begin{document}
\title{Electrostatic and electrokinetic contributions to the elastic
  moduli of a driven membrane}
\author{D. Lacoste\inst{1}
\and G. I. Menon\inst{2} \and M. Z. Bazant\inst{1,3} \and J. F.
Joanny\inst{4}
}                     
%
%
\institute{Laboratoire de Physico-Chimie Th\'eorique, UMR 7083,
ESPCI, 10 rue Vauquelin, 75231 Paris Cedex 05, France \and The
Institute of Mathematical Sciences, C.I.T Campus, Taramani,
Chennai 600 113, India \and Department of Mathematics,
Massachusetts Institute of Technology, Cambridge Massachusetts
02139, USA \and Institut Curie, UMR 168, 26 rue d'Ulm 75005 Paris,
France}
\date{\today}
%

\abstract{ We discuss the electrostatic contribution to the elastic
  moduli of a cell or artificial membrane placed in an electrolyte and
  driven by a DC electric field. The field drives ion currents across
  the membrane, through specific channels, pumps or natural pores. In
  steady state, charges accumulate in the Debye layers close to the
  membrane, modifying the membrane elastic moduli. We first study a
  model of a membrane of zero thickness, later generalizing this
  treatment to allow for a finite thickness and finite dielectric
  constant. Our results clarify and extend the results presented in
  [D.\ Lacoste, M.\ Cosentino Lagomarsino, and J.\ F.\ Joanny,
  Europhys. Lett., {\bf 77}, 18006 (2007)], by providing a physical
  explanation for a destabilizing term proportional to $\kps^3$ in the
  fluctuation spectrum, which we relate to a nonlinear ($E^2$)
  electro-kinetic effect called induced-charge electro-osmosis
  (ICEO). Recent studies of ICEO have focused on electrodes and
  polarizable particles, where an applied bulk field is perturbed by
  capacitive charging of the double layer and drives flow along the
  field axis toward surface protrusions; in contrast, we predict
  ``reverse'' ICEO flows around driven membranes, due to
  curvature-induced tangential fields within a non-equilibrium double
  layer, which hydrodynamically enhance protrusions. We also consider
  the effect of incorporating the dynamics of a spatially dependent
  concentration field for the ion channels.  }
\maketitle

\section{Introduction}

Phospholipid molecules self-assemble into a variety of structures,
including bilayer membranes, when placed in an aqueous
environment\cite{seifert}. The physical properties of such
membranes, at thermal equilibrium, are controlled by a small
number of parameters, including the surface tension and the
curvature moduli. Understanding how these properties are modified
when the membrane is driven out of equilibrium either by
externally applied or internally generated electric fields, is a
problem of considerable importance to the physics of living cells.

Applied electric fields can be used to drive shape
changes in lipid membranes\cite{lipowsky}.  Artificial
lipid vesicles can be produced, via a process called
electroformation, by applying an AC electric field to a
lipid film deposited on an electrode. Applying an electric
field to a vesicle can also lead to the formation of pores
via electroporation, a technique of relevance to gene or
drug delivery. The role of the field in this case is to
introduce transient pores, temporarily removing the barrier
presented by the cell membrane to transmembrane transport.

Large electric fields are also generated internally in living
cells. The transmembrane potential {\it in vivo} results from the
action of a large number of membrane-bound ion pumps and channels.
Resting potentials, and their modulation through excitation, are
crucial to many cell functions \cite{hille}.  Changes in the
transmembrane potential and in the ion charge distribution close
to the membrane accompany shape changes of cell membranes, such as
those which occur when a cell divides. They also provide a means
of communication between cells, as in the classic example of the
action potential of neural cells \cite{kandel,yeung}.

Many aspects of electroformation, electroporation, and of the
collective behavior of ion channels are as yet poorly understood
\cite{lipowsky,Lecuyer}.  This is because most studies of
electrostatic effects in biological membranes have examined
fluctuations at and close to thermal equilibrium
\cite{andelman,pincus,chou,duplantier,helfrich,lekkerkerker,joanny,kleinert}.
However, membranes bearing ion pumps or channels which are driven
by ATP hydrolysis (``active membranes''), or exposed to electric
fields which lead to transmembrane currents in steady state,
cannot be described in terms of equilibrium physics, in the first
case because a non-equilibrium chemical potential for ATP molecules
must be maintained externally to produce such driving and in the
second because a net current cannot flow in any system constrained
by detailed balance.

To proceed beyond an equilibrium description of the membrane, it
is necessary to account for  forces generated by inclusions such
as ion channels, pumps, or artificial pores \cite{PB,RTP,madan}.
An example of such an active membrane was discussed in
Refs.~\cite{PRL_JBManneville,PRE_JBManneville}. In the
experimental work described in these papers, a giant unilamellar
vesicle was rendered active through the inclusion of
light-activated bacteriorhodopsin pumps. These pumps transfer
protons unidirectionally across the membrane as a consequence of
conformational changes, when excited by light of a specific
wavelength. In Ref.~\cite{PRE_JBManneville}, a hydrodynamic theory
for the non-equilibrium fluctuations of the membrane induced by
the activity of the pumps was also developed. This work has
stimulated substantial theoretical interest in the general problem
of a proper description of non-equilibrium effects associated with
protein conformational changes
\cite{gautam,andy,chen,mouritsen,lomholt}.

A major limitation of existing active membrane models is that they
do not describe electrostatic effects associated with ion
transport in detail. These effects are now understood to be very
significant in the biological context.  A recent paper, authored
by two of us~\cite{lacoste}, addressed this limitation by studying
the fluctuations of a membrane containing inclusions such as ion
channels or pumps. Our analysis was based on the use of
electrokinetic equations~\cite{EK,ajdari,iceo} supplemented by a
simple description of ion transport in ion channels.

This paper augments Ref.~\cite{lacoste} by providing details of
the calculations and results presented there.  It also presents
fresh insights into the physical content of some of these results,
while incorporating several new features, as detailed below.  Our
theoretical description of charge fluctuations near the membrane
is in the same spirit as earlier work which examined the stability
of shape fluctuations of a charged membrane using linear
analysis~\cite{kumaran}.  We provide a simple physical picture for
understanding the electrostatically induced part of the surface
tension, which corresponds to a term proportional to $\kps^2$ in
the free energy of the membrane. We do this by relating the
surface tension to an integral over components of the
electrostatic (Maxwell) stresses acting on the membrane and the
fluid in the non-equilibrium steady state.

We also propose a physical interpretation of the term proportional to
$\kps^3$ in the effective free energy, obtained first in
Ref.~\cite{lacoste}. We show that such a term is related to a
nonlinear electrokinetic effect called ``induced-charge
electro-osmosis'' (ICEO)~\cite{iceo}, first described in the Russian
colloids literature~\cite{murtsovkin} and now studied extensively
in microfluidics, since the discovery of electro-osmotic flows over
electrode arrays applying AC voltages~\cite{ramos,ajdari}. Steady ICEO
flows also occur in DC fields around polarizable
metallic~\cite{levitan,harnett} or dielectric~\cite{thamida,yossifon}
surfaces, and broken symmetries generally lead to fluid pumping or
motion of freely suspended polarizable
objects~\cite{iceo,gangwal}. These phenomena are very general and should
also be present in the case of a fluctuating membrane containing ion
pumps and channels.

We also analyze the relaxation of a concentration field
describing a non-uniform, but slowly varying, distribution
of pumps and channels. We include the dynamics of the
concentration field of the channels as in previous studies
of fluctuations of membranes containing active or passive
inclusions\cite{gautam,misbah}. We first study the case
of a membrane of zero thickness. We then generalize the
model to the case of a bilayer of finite thickness and a
finite dielectric constant, but with a uniform distribution
of pumps and channels. This model allows a discussion of
capacitive effects.

The results we present confirm the importance of
capacitive effects in determining electrostatic and
electrokinetic contributions to the elastic moduli of
driven membranes\cite{lacoste}. They can be compared
to results obtained in a recent study of electrostatic
contributions to the elastic moduli of an equilibrium
membrane of finite thickness\cite{lomholt2}. The study
of Ref.~\cite{lomholt2},
which ignores ion transport, predicts a dependance of
the bending modulus and tension as a function of the salt
concentration which we compare to the one obtained in this
paper, in the limiting case where no ion transport occurs
in our model.

Our study is limited to the linear response of the ion
channels and pumps. Real channels have a non-linear
response which is essential for action potentials. Our
study thus excludes these effects as well as other effects
such as electro-osmotic instabilities \cite{leonetti},
which originate in the non-linear response of the ion
channels.

The outline of this paper is the following. In Section~\ref{zero
thickness}, we study a membrane with zero thickness in the linear
response regime. We perform a systematic expansion about a flat
membrane with a uniform distribution of pumps. We then discuss the
charge fluctuations in the Stokes limit. In
Section~\ref{sec:flow}, we analyze ICEO flows around the driven
membrane, emphasizing the basic physics of this new nonlinear
electrokinetic phenomenon. In Section~\ref{sec: concentration
fluct}, we discuss the extension of the model to the case where
the distribution of pumps/channels is non-uniform. In
Section~\ref{finite thickness} we account for the finite thickness
of the membrane. Finally, in Section ~\ref{conclusion} we
summarize the results of this paper and indicate possible
directions for further research. Appendix A describes a mapping
between a driven membrane of finite thickness and an equivalent
zero thickness membrane with appropriately modified boundary
conditions while Appendix B illustrates the solution of the
Stokes equation for the case of the membrane with zero thickness.

\section{Electrostatically driven membrane of zero thickness} \label{zero
thickness}

\begin{figure} {\par {
\rotatebox{0}{\includegraphics[scale=1]{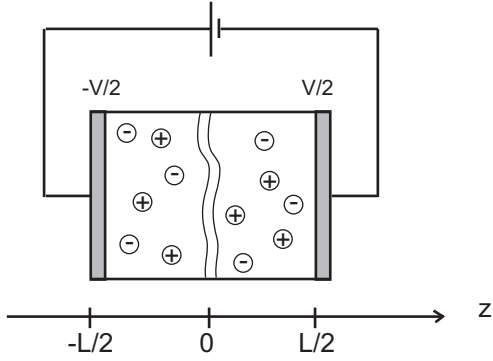}} }
\par}
\caption{Schematic of a quasi-planar membrane embedded in a
symmetric electrolyte. The (bilayer) membrane is represented by
the two wiggly lines near the plane $z=0$. A voltage $V$ is
applied far from the membrane on electrodes separated by a
distance $L$. Note that the electrode at potential $+V/2$ is
called the anode and the one at $-V/2$ the cathode.}
\label{fig:sketch}
\end{figure}

We begin by deriving the equation of motion of a driven membrane
in an electrolyte in the limit in which the membrane has
vanishing thickness and zero
dielectric constant. We work in a linear regime and
consider only steady state solutions. The quasi-planar membrane
is located in the plane $z=0$. It is embedded in an electrolyte
and carries channels for two species of monovalent ions. The
membrane itself is neutral, {\it i.e.} it bears no fixed charge.
There is an imposed potential difference $V$ across the system of
length $L$ as shown in Figure~\ref{fig:sketch}.

The concentrations of the two ions are denoted by $c_k$, where the
index $k$ is 1 for the positive ion ($z_1=1$) and 2 for the
negative ion ($z_2=-1$). A point on the membrane is parameterized,
in a Monge representation valid for small undulations, by a height
function $h(\rp)$, with $\rp$ a two-dimensional vector.

The calculation proceeds via a perturbation theory about the
planar or base state, to first order in the membrane height $h(\rp)$,
assuming a uniform concentration field for the channels/pumps. We
denote dimensionful variables with a superscript $*$, dropping
this superscript for variables which are made dimensionless. A summary of the
dimensionful and dimensionless variables used in this paper
and the correspondence between them is given in Table \ref{table-units}.
To lighten the notation, the inverse Debye length
$\kappa$, the diffusion coefficients for both species
$D_k$, the electrolyte dielectric constant $\epsilon$,
the membrane dielectric constant in the finite thickness
case $\epsilon_m$, the charge of the electron $e$ and the
thermal energy $k_B T$, although dimensionful, will not
carry a superscript $*$.

The potential obeys the Poisson equation
\begin{equation}\label{Poisson non normalized}
\nabla^{*2} \psi^*=- \left( \frac{e c_1^*}{\epsilon} - \frac{e
c_2^*}{\epsilon} \right),
\end{equation}
which becomes
\begin{equation}\label{Poisson}
\nabla^{2} \psi =- \left( \frac{c_1-c_2}{2} \right),
\end{equation}
when the following nondimensional variables are introduced:
$c_k=c_k^*/n^*$, $\psi=e\psi^*/k_B T$, and $x=\kappa x^*$. Here
$n^*$ is the bulk concentration of the electrolyte at large
distance from the membrane and $\kappa$ is the inverse
Debye-H\"{u}ckel length with $\kappa^2=2 e^2 n^* / \epsilon k_B
T$.

We assume a symmetric distribution of ion concentrations on both
sides of the membrane so that the Debye length is the same on both
sides (the asymmetric distribution is discussed in
Ref~\cite{lacoste}).

We  work with dimensionless currents, obtained by introducing
$J_1=J_1^*/(D_1 n^* \kappa)$ and $J_2=J_2^*/(D_2 n^* \kappa)$,
where $D_1$ and $D_2$ are the bulk diffusion coefficient of the
positive and negative ions, and $n^*$ is the bulk concentration of
the electrolyte at large distance away from the membrane.

We use a Poisson-Nernst-Planck approach \cite{hille}, in which ion
currents are treated as constant. Assuming a steady state for ion
concentrations, the equations of charge conservation take the form
\begin{eqnarray}\label{charge
cons} \nabla
 \cdot \left( \nabla c_1 + c_1 \nabla \psi \right) & = & 0, \\
\nabla \cdot \left( \nabla c_2 - c_2 \nabla \psi \right) & = & 0
\label{charge cons2}.
\end{eqnarray}

The non-linear coupling between charge densities and potentials
implies that
general solutions of equations~(\ref{Poisson}-\ref{charge cons2}) are
difficult to obtain analytically. However, as shown in
Ref.~\cite{kumaran},  a solution can be obtained in terms
of a series expansion. In this paper, we  retain only the first
term in such a series expansion. This is the Debye-H\"{u}ckel
approximation, and corresponds to linearizing equations~(\ref{charge
cons}-\ref{charge cons2}). With the definitions $c_1=1+\delta c_1$,
$c_2=1+\delta c_2$, we obtain
\begin{eqnarray}\label{charge
cons lin} \nabla \cdot \left(
 \nabla \delta c_1 + \nabla \psi \right) & = & 0, \\
\nabla \cdot \left( \nabla \delta c_2 - \nabla \psi \right) & = &
0. \label{charge cons lin2}
\end{eqnarray}

\begin{table}
\begin{tabular}{|c|c|c|c|}
\hline
Unit & (1) & (2) & Relation \\
\hline
concentration & $c_k^* $ & $c_k$ & $c_k=c_k^*/n^*$ \\
\hline
electrostatic potential & $\psi^*$ & $\psi$  & $\psi=e \psi^* /(k_B T)$  \\
\hline
length & $x^*$ & $x$  & $x=\kappa x^*$  \\
\hline
particle current & $J_k^*$ & $J_k$  & $J_k=J_k^*/(D_k n^* \kappa)$  \\
\hline
chemical potential & $\mu^*$ & $\mu$  & $\mu=\mu^*/(k_B T)$  \\
\hline
ionic current & $i_k^*$ & $i_k$  & $i_k=i_k^*/(D_k n^* \kappa)$  \\
\hline
conductance & $G_k^*$ & $G_k$  & $G_k=G_k^* k_B T/(D_k n^* e^2 \kappa)$  \\
\hline
charge density (at $z=0^-$) & $\sigma^*$ & $\sigma$  & $\sigma=e \sigma^* /(\kappa \epsilon k_B T)$  \\
\hline
pressure & $p^*$ & $p$  & $p=p^*/(2n^* k_B T)$  \\
\hline
velocity & $v^*$ & $v$  & $v=v^* \eta^* \kappa/(2n^* k_B T)$  \\
\hline
\end{tabular}
\caption{Relation between dimensionful variables (column 1) generally denoted with the superscript $*$ and corresponding dimensionless
variables (column 2). For notational simplicity, as
discussed in the text, the inverse Debye length $\kappa$, the
diffusion coefficients for both species $D_k$, the electrolyte
dielectric constant $\epsilon$, the membrane dielectric constant
in the finite thickness case $\epsilon_m$, the charge of the
electron $e$ and the thermal energy $k_B T$, although
dimensionful, will not carry a superscript $*$.
\label{table-units}}
\end{table}

\subsection{Base state charge distribution}
The base state is defined with respect to the flat membrane, for
which concentration and potential variations can only occur in the
$z$ direction. Denote by $\delta N_1$, $\delta N_2$ and $\Psi$,
the base-state ion concentration profiles and the electrostatic
potential, corresponding respectively to the variables $\delta
c_1$, $\delta c_2$ and $\psi$ of the previous section. Since the
system is driven by the application of an electric field, this
base state is a non-equilibrium steady state. There are constant
particle currents for ions 1 and 2, denoted by $J_1$ and $J_2$,
along the $z$ direction. The equations of charge conservation in
the bulk of the electrolyte are
\begin{eqnarray}\label{base state}
\partial_z \delta N_1 + \partial_z \Psi & = & -J_1,  \\
\partial_z \delta N_2 - \partial_z \Psi & = & -J_2. \label{base state2}
\end{eqnarray}
To simplify notation, we introduce \bb \label{Q} Q= \frac{1}{2}
\left( \delta N_1-\delta N_2 \right). \en Thus, $Q$ represents half the charge
distribution. From equation~(\ref{Poisson}),
we have \bb
\partial^2_z \Psi + Q =0. \label{Poisson normalized} \en

Equations~(\ref{base state}-\ref{Poisson normalized}) are to be solved
with the following boundary conditions:
\begin{equation}\label{BC1b}
\delta N_1(z\rightarrow \pm L/2)=\delta N_2(z\rightarrow \pm
L/2)=0,
\end{equation}
\begin{equation}\label{BC2b}
 \Psi(\pm L/2)=\pm V/2,
\end{equation}
far from the membrane. At the membrane surface, we enforce
continuity of the electric field,
\begin{equation}
\partial_z \Psi_{z=0^+}=\partial_z \Psi_{z=0^-},
\end{equation}
since we assume that the membrane has zero fixed charge.

There is, in general, a discontinuity in the potential, due to electrochemical
equilibrium across the ion channels. This implies a distribution
of surface dipoles on the membrane~\cite{jackson}. In the
appendix, we derive a general Robin-type boundary condition for a
thin dielectric membrane of thickness $d$
\begin{equation}
  \delta_m\, \partial_z \Psi_{z=0^\pm} = \Psi(z=0^+) - \Psi(z=0^-)  \label{eq:sternbc}
\end{equation}
where
\begin{equation}
  \delta_m = \frac{\epsilon \kappa d^*}{\epsilon_m}, \label{eq:deltam}
\end{equation}
which is also used to describe Stern layers and dielectric
coatings on electrodes~\cite{bazant2005}.

The limits of small thickness or small $\epsilon_m$ correspond to
two distinct regimes with either $\delta_m \gg 1$ or $\delta_m \ll
1$. The regime $\delta_m \gg 1$, for equilibrium membranes, is
called the decoupled limit in Ref.~\cite{andelman}, because the
electrical coupling between the layers is suppressed at large
$\delta_m$. This regime typically corresponds to the physical
situation for biological membranes, since $\delta_m \gg 1$ implies
that $\kappa d^* \gg 1/40$. Since the thickness of a typical lipid
bilayer membrane is around $5$nm, this translates to the
requirement that the Debye length $\kappa^{-1} \ll 40d^* \simeq
200$nm, a condition which is usually satisfied. It is thus
tempting to assume that we can take $\delta_m \rightarrow \infty$,
thus reducing the  Robin-type boundary condition to the form
\begin{equation}\label{BC3b}
\partial_z \Psi_{z=0^\pm} = 0.
\end{equation}
The boundary condition of Eq. ~(\ref{BC3b}), equivalent to the
field vanishing at the surface of the membrane, is simple and
convenient to work with for calculational purposes.
However, the precise way in which the decoupled limit should be
approached is, however, somewhat subtle in the non-equilibrium
case.

As  we show quantitatively in  the appendix and discuss
qualitatively further below, in a  calculation in which the zero thickness
case is derived explicitly as a limiting case of the finite thickness problem,
the  $\delta_m \rightarrow \infty$ limit corresponds to unrealistically
large values of the ion channel conductance in comparison to the
biological situation.  This has specific implications for the
sign of the diffuse charge at the membrane surface.
In the first part of this paper, we will nevertheless
assume $\partial_z \Psi_{z=0^\pm} = 0$ for the
following reasons: The use of the
simpler boundary condition of Eq.~(\ref{BC3b}) leads to
considerable calculational simplification as well as
reproduces the profile of the electrostatic potential
to reasonable accuracy. Thus, the physical underpinnings
of many of our results, including the structure of ICEO
flows, can be explained more easily in this limit.
Our results in this limit may
be more relevant to artificial membrane systems
containing pumps and channels or their analogs in which
conductances can be tuned to larger values than attainable
{\em in vivo}. The biologically more relevant general case of
finite-thickness membranes, for which no such simplifying approximation
is made,  is analyzed in the last part of the paper. The boundary condition
(\ref{eq:sternbc}) with finite $\delta_m$ is discussed in
Appendix A.

\begin{figure}[htbp]
  \centering
\rotatebox{0}{\includegraphics[scale=0.7]{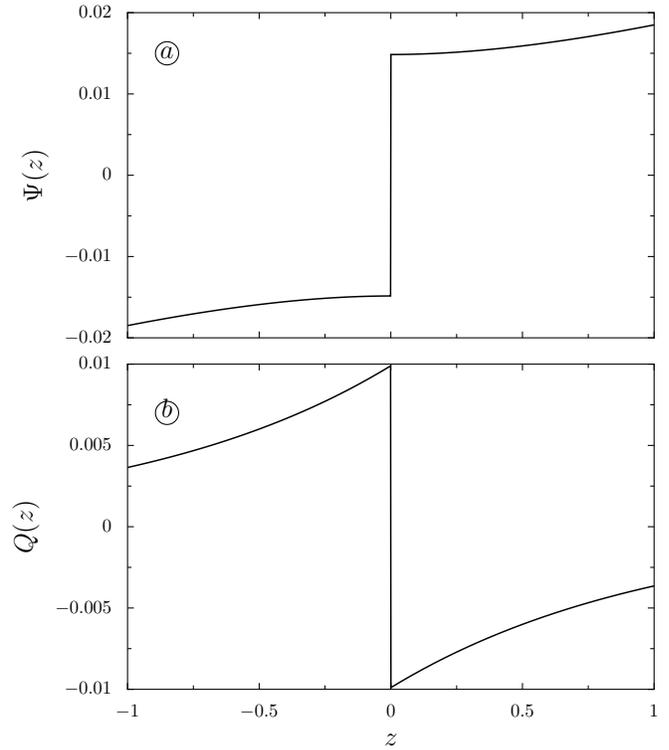}}
\caption{Solutions of the electrokinetic equations for a
    membrane of zero thickness and symmetric ions concentrations.
The electrostatic potential $\Psi(z)$ is shown in (a) and the
quantity $Q(z)$ (which represents half the charge distribution) is
shown in (b).
  We use dimensionless units and the following parameter values
    $V=1$, $L=100$, $G_1=G_2=1$ and $\kps=1$. In these conditions, $\sigma=0.01$.}
 \label{fig:potential zero thickness}
\end{figure}

With the assumptions above, in the limit $L \gg 1$, and for $z>0$, we obtain
\begin{eqnarray}\label{order 0 positive sol}
Q^+(z)  =  -\sigma e^{-z}, \\
\Psi^+(z)=  \sigma \left( z-\frac{L}{2} + e^{-z}
\right)+\frac{V}{2},
\end{eqnarray}
and for $z<0$,
\begin{eqnarray}
Q^-(z)  =  \sigma e^{z}, \\
\Psi^-(z)=  \sigma \left( z+\frac{L}{2} - e^{z}
\right)-\frac{V}{2}, \label{order 0 negative sol}
\end{eqnarray}
where \bb \sigma=\frac{1}{2} \left( -J_1+J_2 \right),
\label{def_sigma} \en is the normalized electrical current, and
the superscripts $\pm$ refer to the regions of $z>0$ and $z<0$
respectively. The electric field component along $z$ is
$E_z^\pm(z)=-\partial_z \Psi^\pm$. For $z>0$, we thus have
\begin{equation}\label{E+}
E_z^+(z)=\sigma \left( e^{-z}-1 \right),
\end{equation}
and for $z<0$
\begin{equation}\label{E-}
E_z^-(z)=\sigma \left( e^{z}-1 \right).
\end{equation}
%
Note that in our dimensionless formulation $\mp \sigma = Q^+(0^\pm)-Q^-(0^\pm)$ is also the normalized diffuse (ionic) charge density evaluated at the membrane surfaces, $z=0^\pm$.
The potential and the charge distribution calculated here are shown in Fig.~\ref{fig:potential zero thickness}.

In this model, the diffuse layers are intrinsically out of equilibrium
and the non-zero DC current influences the distribution of ions
through $(\ref{def_sigma})$. Note that the sign of the non-equilibrium
diffuse charge is negative on the positive side of the membrane {\it
  i.e.} $z=0^+$, which we call the cathodic side (although it faces
the anode) since positive charge flows towards it. We remind the
reader that the cathode is the electrode located at $z=-L/2$ (see
Figure~\ref{fig:sketch}), towards which positively charged cations
drift, while negatively charged anions drift toward the anode at
$z=L/2$.

This sign of the diffuse charge is unexpected -- it
is opposite to what is found in standard models for
electrodes in a galvanic cell~\cite{bazant2005} or
(potentiostatic) electrodialysis membranes~\cite{zaltzman}
or in other related models of a membrane in an electric
field \cite{pierre}, where diffuse charge resides in
thin layers in Boltzmann equilibrium (up  to the limiting
current) and has the opposite sign, positive at the
cathodic and negative at the anodic surfaces. Since
biological membranes are typically much less conductive
than the surrounding electrolyte, it is intuitively reasonable
that positive charges should pile up under the action of
the electric field directed from the anode to the cathode,
near the positive side of the membrane. The "wrong" sign of
the charge distribution
obtained in Eqs.~\ref{order 0 positive sol}-\ref{order
0 negative sol} and shown in Fig.~\ref{fig:potential
zero thickness} is thus an artefact of the approximation of
zero thickness and zero dielectric constant. Physically,
this unusual behavior may be attributed to the following:
the positive charges which should pile up
near the positive side are overcompensated
by a charge of the opposite sign, in order to satisfy the boundary
condition Eq.~\ref{BC3b} of a zero electric field on
the membrane.

Taking the limit of the general Robin-type boundary condition
makes  sense if $1/\delta_m$ vanishes. In reality, however, $\delta_m$
is finite and although it is larger than one, it is
incorrect to assume an infinite $\delta_m$  in the
calculation of the charge distribution. Using the more
general boundary condition (\ref{eq:sternbc}) with finite
$\delta_m$ derived in the appendix A, and which
is appropriate to describe a membrane of finite thickness
and finite dielectric constant, we show in section 5 of
this paper that {\em both signs of the charge distribution
are possible in principle}. Under normal biological conditions,
as we demonstrate using numerical estimates at the beginning of
section 5.2, the membrane is much less conductive than
the surrounding electrolyte and the diffuse charge
distribution has the opposite sign as compared to that of
Fig.~\ref{fig:potential zero thickness}.

We now discuss the boundary condition for the ion current
at the membrane surface.
This is ensured by choosing a specific relation between the
current and the voltage at an ion pump or channel. In general such
a relation is non-linear. We assume, for simplicity, a linear
relation \bb J_k=-G_k \Delta \mu_k, \label{Eq: BC for currents}
\en where $\Delta \mu_k$ is the normalized chemical potential
difference of ion $k$ across the membrane, and $G_k$ is a
normalized conductance. This (dimensionless) conductance $G_k$ is
related to the dimensionful conductance per unit area $G_k^*$ by
\bb G_k=\frac{G_k^* k_B T}{D_k^* n^* e^2 \kappa},
\label{conductance} \en where the normalizing factor represents
the conductance per unit area of a layer of electrolyte of
thickness equal to $1/\kappa$ (one Debye layer). The normalized
chemical potentials are defined by
\begin{eqnarray} \label{def of mu}
\mu_1 & = & \delta N_1+\Psi, \\
\mu_2 & = & \delta N_2-\Psi, \end{eqnarray} and \bb \Delta
\mu_k=\mu_k(z=0^+)-\mu_k(z=0^-). \en

The currents are now determined self-consistently as
\begin{eqnarray}\label{final currents}
J_1 & = & -\frac{G_1 V}{1+ G_1 L}, \\
J_2 & = & \frac{G_2 V}{1 + G_2 L}. \label{final currents2}
\end{eqnarray}
Restoring dimensions, the electrical current is \cite{lacoste} \bb
i_k^*= \frac{-G_k^* v_k^*}{1+ \frac{G_k^* L^* k_B T}{D_k n^*
e^2}}, \label{exp current} \en with $i_k^*=z_k e J_k^*$ the part
of the total electric current associated with ion $k$ of charge
$z_k=\pm 1$, $v_k^*=V^*-V_{Nernst,k}^*$, and $V_{Nernst,k}^*$ the
Nernst potential of ion $k$, which is zero here due to our
assumption of symmetric concentrations. Note that $\sigma^*$ has
the units of charge per unit surface and is the surface charge of
the Debye layers. It is related to $\sigma$ defined in
equation~(\ref{def_sigma}) by \bb \sigma^*=\frac{k_B T \kappa
\epsilon \sigma}{e}. \label{sigma dim ful-less} \en The equivalent
of equation~(\ref{def_sigma}) in dimensionful form is \bb
\sigma^*=-\frac{1}{\kappa^2} \left( \frac{i_1^*}{D_1} +
\frac{i_2^*}{D_2} \right). \label{sigma_with dimension} \en This
equation expresses the conservation of charge inside the Debye
layers: for each ion $k$, the contribution in the surface charge
of the Debye layer $\sigma^*$, is the product of the total
electric current per unit area $i_k^*$ carried by ion $k$, with
the diffusion time $1/\kappa^2 D_k$ for the ion to diffuse over a
length scale equal to the Debye length.

Equation ~(\ref{exp current}) is consistent with the usual electric
representation of ion channels in the ohmic regime in which
the contribution of each ion taken in parallel. There
are two conductances for each ion, accounting for the
contributions of the electrolyte on both sides, and
an electromotive force $E_k$ in series\cite{lacoste}.
The form of equations~(\ref{order 0 positive sol}-\ref{order 0
negative sol}) is general and holds even when a non-linear
current versus chemical potential relation is used in
place of equations~(\ref{Eq: BC for currents}). However, our
approach will be restricted to the linear regime for the
ion channel response.

We stress that the form of this base state is general in the sense
that the precise origin of the ion
currents is immaterial because these currents are constant (independent of
$z$). A qualitatively similar base state would describe the situation
where such currents are created {\it internally} by active pumps,
in the absence of any externally imposed potential difference or
concentration gradients.

To complete the characterization of the base state, we calculate
the stresses on the membrane. We define the stress tensor by \bb
\tau_{ij}^*=\tau_{ij}^{H*}+ \tau_{ij}^{M*}, \en where
$\tau_{ij}^{H*}$ and $\tau_{ij}^{M*}$ are the hydrodynamic and
Maxwell stress tensors respectively, defined by \bb
\tau_{ij}^{H*}=-P^* \delta_{ij} + \eta^* \left(
\partial_i^* v_j^* +
\partial_j^* v_i^* \right), \label{tau_H*}
\en where $\eta^*$ is the solvent viscosity and \bb
\tau_{ij}^{M*}=\epsilon \left( E_i^* E_j^* - \delta_{ij} {E^*}^2/2
\right). \label{tau_Maxwell*} \en In dimensionless form these are
\bb \tau_{ij}=\tau_{ij}^H+ \tau_{ij}^M, \en with \bb
\tau_{ij}^H=-P \delta_{ij} +\left(\partial_i v_j +
\partial_j v_i \right), \label{tau_H}
\en and \bb \tau_{ij}^{M}=E_i E_j - \delta_{ij} E^2/2,
\label{tau_Maxwell} \en where $E_i$ is the ith component of the
electric field. The pressure $P$ is the osmotic pressure of the
ions in the Debye layers. In the base state, the condition $\nabla
\cdot \tau=0$ is equivalent to $\nabla P=Q E$, with $Q$ and $E$
given by equations~(\ref{order 0 positive sol}-\ref{order 0 negative
sol}).

With the boundary condition $P(z=\infty)=0$, we obtain
\begin{equation}\label{Pressure}
P(z)=\sigma^2 \left( -e^{-z}+ \frac{e^{-2 z}}{2} \right),
\end{equation} for $z>0$. Using
equations~(\ref{tau_H}-\ref{tau_Maxwell}), the  stress on the
positive side is calculated as $\tau_{zz}^+(z=0)=\sigma^2/2$. It is
straightforward to check that the same contribution exists on the
negative side.  Thus, overall, normal stresses are balanced in the base state,
although a pressure gradient is present.

\subsection{Interpretation of the electrostatic contribution to
the surface tension} 
\begin{figure} {\par {
\rotatebox{0}{\includegraphics[scale=0.8]{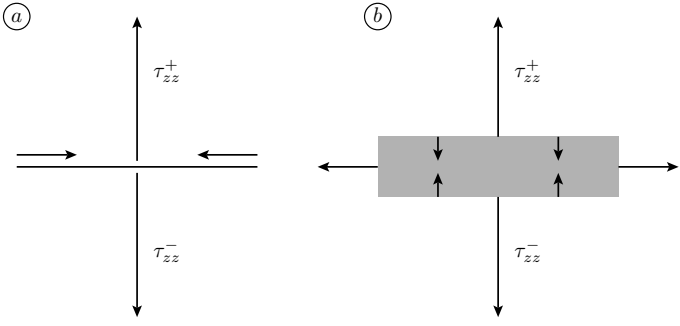}} }
\par}
\caption{Electrostatic corrections to the membrane tension for a
membrane of zero thickness (a) and finite thickness (b). The
Maxwell stresses are represented by vertical arrows while the
horizontal arrows represent the resulting tension on the membrane
as a consequence of incompressibility.} \label{fig:tension}
\end{figure}

The extensive normal stresses discussed in
the previous subsection can be argued to result in a positive
electrostatic correction to the membrane surface tension (see
Figure~\ref{fig:tension}). This correction to the membrane tension
$\Sigma$ can be obtained from the knowledge of the electric field
in the base state $E^{(0)}$ \cite{lacoste}. In our geometry, this
correction can be written as \bb \label{surfacetension0} \Sigma=
\int_{-\infty}^{\infty} \left[ \tau_{xx}(z) - \tau_{zz}(z) \right]
dz, \en where $\tau_{xx}$ and $\tau_{zz}$ are components of the
stress tensor. This derivation assumes
incompressibility\cite{widom}.

The electrostatic contribution to the surface tension is obtained from
the Maxwell stress by $\Sigma= \Sigma_0+\Sigma_1$ with : \bb
\label{surfacetension} \Sigma_0=-\int_{-L/2}^{L/2}
(E_z^{(0)})^2(z) dz, \en and \bb \label{surfacetensionb}
\Sigma_1=\frac{L}{2} \left[
  (E_z^{(0)})^2(z \rightarrow \infty) + (E_z^{(0)})^2(z \rightarrow -\infty)
  \right]. \en
The term in $\Sigma_1$ ensures that the stress tensor remains
divergence free. Both $\Sigma_0$ and $\Sigma_1$  contain
contributions proportional to $L$, which originate from the
pressure gradient in the fluid. As expected, these
terms cancel each other in $\Sigma$. Substituting our previous
expression for the electrostatic potential into
equation~(\ref{surfacetension}), we find that $\Sigma=3 \sigma^2$.
We will recover this result in the next section using a different
method.

We now illustrate our physical picture for the origin of this
electrostatic correction to the membrane tension. As shown in Figure
\ref{fig:tension}, for a membrane of zero thickness, only Debye layers
above and below the membrane contribute to the electrostatic
correction to the membrane tension. The electrostatic force acting on
the induced charges in the Debye layers on the positive and negative
sides creates extensive stresses $\tau_{zz}^\pm$ near the
membrane. These stresses, by incompressibility, tend to reduce the
membrane area, thus producing an increase in the membrane
tension. This can be termed as the "outside" contribution to the
surface tension. In the case of a membrane of finite thickness there
is, in addition to the "outside" contribution, an "inside"
contribution. The "inside" contribution is in general dominant,
because the largest voltage drop in this problem occurs across the
membrane. This is a consequence of the large mismatch in dielectric
constants between the membrane and the electrolyte ($\delta_m\gg 1$).

The "inside" contribution arises from compressive stresses
(represented as opposing arrows within the shaded area on the
figure on the right), which are generically present in any
capacitor. These compressive stresses, directed along the $z$ direction,
produce lateral extensional stresses due to the conservation of
the inside volume of the membrane. These stresses act to increase
the membrane area, thus producing a negative electrostatic correction
to the membrane tension. This contribution has been recognized to
drive instabilities in membranes when a normal DC electric field
is applied \cite{Lecuyer,lomholt2,pierre}.

Recent experimental studies on the fluctuation spectrum
of active membranes containing bacteriorhodopsin exhibit
a lowering of the membrane tension in active vesicles
as compared to passive ones \cite{PRL in preparation}.
This observation is consistent with the interpretation
suggested above, where the lowering of the tension would
be caused by a change in normal Maxwell stresses as a consequence
of ion fluxes in or out of the vesicle.  Although this
interpretation appears plausible, alternate explanations
are possible: further experimental work and theoretical
modeling are necessary to confirm this proposal.

\subsection{Charge fluctuations}
In a linear approximation, the electrostatic potential can be
written as a superposition of the base state contribution
$\psi^{(0)}$ and a contribution linear in the membrane height
field $\psi^{(1)}$. We work in the quasi-static approximation,
which corresponds to angular frequency $\omega^*$ such that $\omega^* \ll
D_k^* \kappa^2$. This approximation means that the membrane
fluctuations occur on a time scale which is much slower than the
time over which the electrostatic configuration adjust itself.
Simple numerical estimates show that there is indeed such a
separation of time scales \cite{lomholt2}. This approximation
allows us to solve the electrostatic problem for a fixed weakly
curved geometry of the membrane.

With our previous notation: $q^{(0)}=Q$, $c_k^{(0)}=N_k$ and
$\psi^{(0)}=\Psi$ in the base state, we now have
\begin{equation}\label{def potential}
\begin{array}{ccc}
\psi(\kp,z) = \Psi(z) + \psi^{(1)}(\kp,z), \\
q(\kp,z) = Q(z) + q^{(1)}(\kp,z), \\
c_1(\kp,z) = N_1(z) +  c_1^{(1)}(\kp,z), \\
c_2(\kp,z) = N_2(z) +  c_2^{(1)}(\kp,z).
\end{array}
\end{equation}
We use the following definition of Fourier
transforms of an arbitrary function $g(\rp,z)$
\begin{equation}\label{def FT}
g(\kp,k_z)=\int d\rp dz e^{-i \left( \kp \cdot \rp + k_z z \right)
} g(\rp,z),
\end{equation}
and the inverse Fourier transform,
\begin{equation}\label{def inverse FT}
g(\rp,z)=\frac{1}{(2 \pi)^3} \int d\kp dk_z e^{i \left( \kp \cdot
\rp + k_z z \right)} g(\kp,k_z).
\end{equation}

Consider now the contribution linear in the membrane height
field $\psi^{(1)}$. The equations for the Fourier transforms of
the charge distribution \bb q^{(1)}(\kp,z)=\frac{1}{2} \left(
c_1^{(1)}(\kp,z)-c_2^{(1)}(\kp,z) \right) \en and of the
electrostatic potential $\psi(\kp,z)$ follow from
equations~(\ref{Poisson}-\ref{charge cons lin2}),
\begin{eqnarray}\label{eq 1st order charge fluctuation}
\left( \partial_z^2 -k_\perp^2 \right) \psi^{(1)}(\kp,z) +q^{(1)}(\kp,z) & = & 0, \\
\left( \partial_z^2 -k_\perp^2 \right) \left(q^{(1)}(\kp,z) +
\psi^{(1)}(\kp,z) \right) & = & 0.
\end{eqnarray}
Since $L$ is much larger than a Debye length, we can take the
boundary conditions far from the membrane to be
$\psi^{(1)}(\kp,\pm \infty)=q^{(1)}(\kp,\pm \infty)=0$.

The relation between the current and the voltage at the membrane
surface incorporating the contribution linear in the membrane
height field is then calculated as
\begin{equation}\label{def potential2}
\begin{array}{ccc}
\partial_z \left( c_1^{(1)}(\kp,z)+\psi^{(1)}(\kp,z) \right)_{z=h(\rp)}= G_1 \times \\
\left( c_1^{(1)}(\kp,0^+)-c_1^{(1)}(\kp,0^-) + \psi^{(1)}(\kp,0^+)- \psi^{(1)}(\kp,0^-) \right), \\
\partial_z \left( c_2^{(1)}(\kp,z)-\psi^{(1)}(\kp,z) \right)_{z=h(\rp)}= G_2 \times \\
\left( c_2^{(1)}(\kp,0^+)-c_2^{(1)}(\kp,0^-) -
\psi^{(1)}(\kp,0^+)+ \psi^{(1)}(\kp,0^-) \right),
\end{array}
\end{equation}
These relations, the boundary conditions for the potential and the
ion concentrations at infinity, as well as equation~(\ref{eq 1st order
charge fluctuation}) are all satisfied when
$c_1^{(1)}(\kp,z)=-c_2^{(1)}(\kp,z)$ and
$q^{(1)}(\kp,z)=-\psi^{(1)}(\kp,z)$. This implies a zero flux
boundary condition for the contribution to first order in the
membrane height field  \bb \left(
\partial_z q^{(1)} +
\partial_z \psi^{(1)} \right)_{z=0}  =0. \label{BC flux simple 10} \en
At this order, whether the fluxes are directed along the normal
$\vn$ rather than along $\vz$ is irrelevant, since the
difference between $\vn$ and $\vz$ only introduces corrections to
equation~(\ref{BC flux simple 10}) which are of higher order than linear
in $h$.

As a consequence, $\psi^{(1)}$ only depends on the zeroth
order solution through the boundary conditions for the potential.
The boundary condition for the total potential corresponds to a
vanishing electric field at the membrane perturbed surface and is
thus \bb \left(
\partial_z \psi \right)_{z=h(\rp)} = \left( h(\rp)
\partial_z^2 \Psi + \partial_z \psi^{(1)} \right)_{z=0}=0.
\en

Our final results for the potential thus are, for $z>0$, \bb
\label{BC potential z>0} \psi^{(1)}(\kp,z)=-q^{(1)}(\kp,z)=
\frac{h(\kp) \sigma}{l} e^{-lz}, \en and for $z<0$, \bb \label{BC
potential z<0} \psi^{(1)}(\kp,z)=-q^{(1)}(\kp,z)= \frac{h(\kp)
\sigma}{l} e^{lz}, \en where we have introduced \bb
l=\sqrt{k_\perp^2+1}, \en the characteristic inverse length of the
electrostatic potential. The charge distribution and potential are
even functions of $z$. Figure \ref{fig:potential zero
thickness} exhibits the potential $\Psi(z)$ and $\psi^{(1)}(\kp,z)$ in
dimensionless units. This illustrates the discontinuity of the
potential across the membrane, and the fact that the electric
field vanishes at the membrane surface at zeroth order, as
imposed by equation~(\ref{BC3b}).

\subsection{Membrane elasticity and force balance}
In order to describe the coupling between the charge fluctuations
in the electrolyte and the membrane, the Stokes equations
must be solved with the appropriate boundary conditions, namely,
the continuity of the velocity and the tangential stress
constraints.

The elastic properties of the membrane are described by an
Helfrich free energy \bb \label{mb free energy1}
F_{mb}=\frac{1}{2} \int d^2 \rp [ \kappa_0 \left( \nabla^2 h
\right)^2 + \sigma_0 \left( \nabla h \right)^2 ], \en where
$\kappa_0$ is the bare bending modulus and $\sigma_0$ is the bare surface
tension of the membrane.

The components of the stress tensor which act normal to the membrane are
discontinuous, and that discontinuity is equal to the restoring
force exerted by the membrane on the fluid, which is equal to
 \bb \label{restoring force}
-\frac{\partial F_{mb}}{\partial h(\rp)}=\sigma_0 \triangle h(\rp)
- \kappa_0 \nabla^4 h(\rp). \en

\subsection{Linear hydrodynamics of the membrane-fluid system}
The equation of motion of the fluid, in the limit of
low-Reynolds number and slow variation with time, is
the Stokes equation supplemented by the condition of
incompressibility.
The governing equations, in dimensionless form, incorporating
an arbitrary force density ${\bf f}$, are
\begin{eqnarray}
\nabla \cdot \vv & = & 0, \label{incompressibility} \\
-\nabla p + \triangle \vv + \vf & = & 0. \label{Stokes}
\end{eqnarray}
We have rescaled the velocity by $2n^* k_B T/\eta^* \kappa$, and
the pressure by $2n^* k_B T$. The Stokes equation
(equation~(\ref{Stokes})) can be written equivalently as $\nabla \cdot
\tau =0$ in terms of the stress tensor of the fluid introduced in
equations~(\ref{tau_H}-\ref{tau_Maxwell}).

In view of the invariance of the problem with respect to
translations parallel to the membrane surface, it is helpful to use the 2D
Fourier representation introduced in Eqs.~\ref{def FT}-\ref{def inverse FT}.
As shown in Refs.~\cite{seifert,thomas bickel}, all vector fields in this problem
can be decomposed into three components: longitudinal ({\it i.e.} along $\kp$), transverse or
normal ({\it i.e.} along $\vz$). These vectors form the
triad $(\kph, \vn, \vt)$, where $\kph=\kp/\kps$,
$\vn=\vz$ and $\vt=\kph \times \vn$. In such a coordinate system,
the incompressibility
condition takes the form
\bb
\partial_z v_z + i \kp \cdot \vp  =  0, \label{incompressibility2}
\en
and the Stokes equations become
\begin{eqnarray} \label{v along perp}
-i \kp p - \kps^2 \vp + \fp + \partial_z^2 \vp & = & 0,  \\ \label{v along z}
- \partial_z p + \partial_z^2 v_z - \kps^2 v_z + f_z & = & 0, \\ \label{v transverse}
\partial_z^2 v_t - \kps^2 v_t + f_t & = & 0. 
\end{eqnarray}

In the bulk of the electrolyte, we know the expression of the force $\vf$. It is
 the electrostatic force acting on the local charge distribution, thus
\begin{equation}\label{force density}
\vf = -q \nabla \psi,
\end{equation}
which at first order in the perturbation defined
in equation~(\ref{def potential}), is
\begin{eqnarray} \label{force components}
\fp (\kp,z) & = & -i \kp \psi^{(1)}(\kp,z) Q(z), \nonumber \\
f_z (\kp,z) & = & -\nabla_z \psi^{(1)}(\kp,z) Q(z) \nonumber \\
            &  - & q^{(1)}(\kp,z) \nabla_z \Psi(z).
\end{eqnarray}

Note that the force $\vf$ above has no components
along the transverse direction, and that the equation for $v_t$ is decoupled from
that of the other components of the velocity. In view of the boundary conditions
appropriate here, we have $v_t=0$ everywhere. Thus, we only need to
consider the longitudinal and normal components. Although these components appear
coupled in Eqs.~\ref{v along perp}-\ref{v along z}, they can in fact be decoupled and the pressure
can be eliminated. Indeed, the pressure can be obtained
from Eq.~\ref{v along perp}. After using the
incompressibility condition, the expression can be written
in terms of only $v_z$ and $\fp$:
\bb
p=-\partial_z v_z + \frac{1}{i \kps} \fp + \frac{1}{\kps^2} \partial_z^3 v_z. \label{pressure}
\en
After inserting this expression for the pressure in Eq.~\ref{v along perp}
and using the incompressibility condition of Eq.~\ref{incompressibility2},
one finds that the normal component of the velocity $v_z$
obeys a single fourth order differential equation
\bb \label{4order}
\left( \partial_z^2 - \kps^2 \right) \left( \partial_z^2 - \kps^2 \right) v_z
+ \left( q^{(1)} \partial_z \Psi -\partial_z \psi^{(1)} Q \right)=0.
\en

The boundary conditions are: (i) continuity of the velocity, (ii)
continuity of tangential constraints and (iii) discontinuity of the
normal-normal component of the stress tensor. The equations of
continuity for  the velocity field are
\begin{eqnarray}
v_z(z=0^+)=v_z(z=0^-) & = & \frac{\partial h(\rp)}{\partial t},
\label{permeation} \\
\vp(z=0^+)=\vp(z=0^-) & = & 0.
 \label{BC for 1st order}
\end{eqnarray}
We have assumed, in writing equation~(\ref{permeation}), that
there is a negligible amount of permeation of water across the
bilayer, an assumption which should be suitable to describe most
ion channels \cite{PRE_JBManneville}. Although the membrane does
permit the two-way flow of ions across it, the mechanical response
of the membrane is dictated primarily by its relatively low
permeability to water. Far from the membrane, we expect that
\bb
v_z(z\rightarrow \pm \infty)=p(z\rightarrow \pm \infty)  =  0. \en

Interestingly, the boundary conditions for
the transverse component of the velocity Eq.~\ref{BC for 1st
order} together with the incompressibility condition
Eq.~\ref{incompressibility} implies another continuity relation for the
derivative of $v_z$ \cite{thomas bickel}:
\bb \label{continuity dvz/dz} \left( \frac{\partial v_z}{\partial z} \right)_{z=0^+}= \left( \frac{\partial v_z}{\partial z}
\right)_{z=0^-}. \en

The boundary conditions expressing the continuity of the
tangential constraints (ii) and the discontinuity of the
normal-normal component of the stress tensor (iii) are
\begin{eqnarray}
-\tau_{\perp z}(z=0^+) + \tau_{\perp
z}(z=0^-) & = & 0, \label{BC stress transverse} \\
 -\tau_{zz}(z=0^+) +
\tau_{zz}(z=0^-) & = & -\frac{\partial F_{mb}}{\partial
h(\rp)}. \label{BC stress normal}
\end{eqnarray}
It is important to stress that this problem cannot be formulated only
in terms of bulk
forces, {\it i.e} of the divergence of a stress tensor, because
the hydrodynamic and Maxwell stress tensors enter the boundary
conditions at the membrane surface explicitly. For this reason,
the force of Eq.~\ref{force density} only holds in the bulk, but the
force localized on the membrane surface is unknown in this problem. It must
be determined by enforcing the velocity and the stress
boundary conditions.

\subsection{Effective elastic moduli of the membrane}
In this section, we give the equation of motion of the membrane which is
obtained from the solution of the linear hydrodynamic equations.
It is convenient to introduce the growth rate $s$ of the
height fluctuation defined by $h(\rp,t)=h(\kp) \exp (i \kp \cdot \rp
+st)$, so that the continuity equation for the normal component of
the fluid velocity Eq.~\ref{permeation}
can be written equivalently as
\bb
v_z(\kp,z=0^\pm)= s h(\kp).
\en
As shown in Appendix B, the following equation of motion for the membrane results
\bb s= -\frac{1}{4} \left( 3
\sigma^2 + \sigma_0 \right) \kps  + \sigma^2 \kps^2 - \left(
\frac{3 \sigma^2}{16}+\frac{\kappa_0}{4} \right) \kps^3.
\label{growth rate} \en
In the particular case where $\sigma=0$, corresponding to the case where there are
no bulk electrostatic forces $\vf=0$, we recover a well-known relation
\cite{PRE_JBManneville}, which can be written
\bb
s= -\frac{1}{4} \sigma_0 \kps   -
\frac{\kappa_0}{4} \kps^3,
\label{growth rate s=0} \en
or equivalently
\bb \label{eq motion 2}
\frac{\partial h(\kp)}{\partial t}= - \frac{1}{4 \kps}
\frac{\partial F_{mb} }{\partial
h(\kp)}. \en

A convenient way to describe the effect of the additional terms arising in
the equation of motion due to the electrostatic force
when $\sigma \neq 0$ is to generalize Eq.~\ref{eq motion 2} to
\bb \label{eq motion 3}
\frac{\partial h(\kp)}{\partial t}= - \frac{1}{4 \kps}
\frac{\partial \left( F_{mb} + \delta
F_{mb} \right) }{\partial
h(\kp)}, \en
where we have introduced an effective free energy
$\delta F_{mb}$ to account for the contribution of electrostatic stresses on
the membrane. We stress that this definition does not
imply that this effective free energy is to be understood in
thermodynamic terms. It is merely a convenient way of
understanding the role of each separate contribution to the stress
tensor arising out of membrane fluctuations.
Writing this effective free energy as \bb \label{mb delta free
energy} \delta F_{mb}=\frac{1}{2} \int d^2 \kp h(\kp) h(-\kp) [ K
\kps^4 + \Sigma \kps^2 + \Gamma \kps^3 ], \en we obtain
electrostatic corrections to the elastic moduli of the membrane.

 Since the second term in the right hand
side of equation (\ref{growth rate}) is positive (a consequence of
the fact that $\Gamma$ is negative), a finite wavelength
instability of a membrane or vesicle of low tension can occur when
$\sigma$ is sufficiently high \cite{lacoste}. We provide an estimate
of the characteristic wavevector $k_c$ below.

When non-thermal noise can be neglected, the fluctuation spectrum
of the membrane height field can be obtained from equations~(\ref{mb
free energy1}-\ref{mb delta free energy}), \bb \langle
|h(\kp)|^2 \rangle = \frac{1}{ \left(\sigma_0 + \Sigma \right)
\kps^2 + \Gamma \kps^3 + (\kappa_0 + K) \kps^4 }.
\label{fluctuation spectrum} \en Such a spectrum is shown in
Figure~\ref{fig:spectrum}.

Our results are the following: We find an
electrostatic correction to the surface tension $\Sigma=3
\sigma^2$, thus recovering the result obtained in
equation~(\ref{surfacetension}). There is also a positive correction to
the bending modulus which is $K=3 \sigma^2/4$. Such terms are not
surprising because they are present with the same sign in
equilibrium charged membranes \cite{andelman}. What is, however,
surprising is the presence of a new purely non-equilibrium term in
factor of $\kps^3$ in the free energy, $\Gamma= -4 \sigma^2$. We
propose a physical interpretation for this term in the
sub-section which follows.

In dimensionful form, these moduli are $\Sigma^*=3
(\sigma^*)^2/\kappa$, $K^*=3 (\sigma^*)^2/4 \kappa^3$ and
$\Gamma^*=-4 (\sigma^*)^2 / \kappa^2$, in terms of $\sigma^*$ the
dimensionful surface charge, in agreement with
Ref.~\cite{lacoste}. For order-of-magnitude estimates, with
$V^*=50$mV, $L^*=1\mu$m, $G_1^*=G_2^*=10 \Omega^{-1}$/m$^2$,
$D_1=D_2=10^{-5}$cm$^2/s$ and
    $n^*=16.6$mM, we obtain  $\Sigma^*= 3.2 \cdot 10^{-16}$ Jm$^{-2}$,
    $\Gamma^* = -10^{-24}$ Jm$^{-1}$
and $K^* = 10^{-13}k_BT$ . Although the ion flux is typical of ion
channels, the moduli $\Sigma^*$, $\Gamma^*$ and $K^*$ are very
small due to the strong dependance of these moduli on
$\kappa^{-1}$, which is only $2.3$nm here. As we show below, these
low values also reflect the fact that we have, until now, neglected
the bilayer character of the membrane and its finite capacitance.

The characteristic wavevector of the finite wavelength instability discussed in Eq.~\ref{growth rate} is $k_c=-\Gamma/2(K+K_0)$ \cite{lacoste}. With the numerical estimates given above,
and a typical value for the bare bending modulus of the membrane $K0$ of 10kT, one finds that $k_c$ is of the order of $10^{-5}$m$^{-1}$. This corresponds to a very large length scale, which indicates that this instability is unlikely to be observed in practice. A very different instability arises in a membrane of finite thickness when the tension becomes negative. That instability is a zero wavelength instability and is a real effect \cite{Lecuyer,lomholt2,pierre}.

\section{ Electro-osmotic flow induced around the membrane }
\label{sec:flow}

\begin{figure}
{\par { \rotatebox{0}{\includegraphics[scale=0.9]{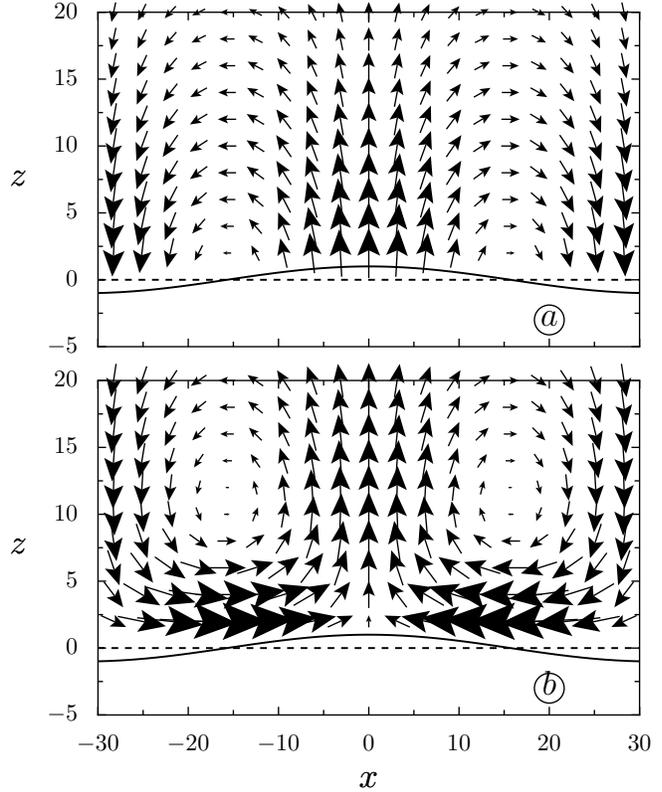}} } \par}
  \caption{ Two types of fluid flow around a perturbed, driven
    membrane. (a) The membrane bending mode with associated flow
    field \cite{levine}, in the absence of any applied electric field: $\sigma=0$.
     In this case, it is the motion of the membrane which drives the flow
     field by the incompressibility condition. For a growing
    sinusoidal perturbation, streamlines connect the peaks to
    the valleys.
    The membrane seen edge-on as a solid line, undergoes a bending
    wave of wavevector 0.1 (or of wavelength $20\pi$) and of amplitude 0.1.
    Note that the height of the membrane has been multiplied by an
    extra factor 10 for improved visualization.
    The undeformed
    membrane is shown edge-on as a dashed line.
    (b) Vortices of induced-charge electro-osmosis (ICEO)
    for a non-moving curved membrane, due to effective slip from the
    valleys to the peaks, as explained below. Unlike the case (a),
    here it is the flow field induced by ICEO which determines the
    modulation of this non-moving membrane.
    An applied field is applied which corresponds to $\sigma=30$.
    In this example, as in
    biological membranes, the double layers are thin compared to the
    wavelength of the perturbation $\kps = \kps^* / \kappa \ll 1$. The
    calculation also assumes linear response to a small amplitude
    perturbation, $\kappa h^* \ll 1$. In these figures,
     this condition is satisfied since
    $h=0.1$, and the unit length corresponds to one Debye length.}
\label{fig:thinDLflow} \label{fig:hydro flows}
\end{figure}

In this section, we propose an interpretation of the cubic term in
$\kps$ with coefficient $\Gamma$ in the effective free energy
obtained above. Our arguments are based on the existence of
nonlinear electro-osmotic flow around a curved membrane. This
direct electrokinetic effect is present in addition to the usual
viscous flow caused by membrane motion~\cite{levine}, shown in
Figure~\ref{fig:thinDLflow}(a), which couples indirectly to the
electric field.

In our geometry, the electric field is directed mainly along the
$z$ direction. Our analysis of charge fluctuations indicated that
perturbations in the membrane shape induced a tangential component
of the electric field near the membrane. Such a tangential
electric field acts on the diffuse charge in the diffuse layers,
creating a effective hydrodynamic slip relative to the
instantaneous membrane position. This electrokinetic effect
creates an array of counter-rotating vortices around the membrane,
illustrated in Figure~\ref{fig:thinDLflow}(b), which tend to
enhance shape perturbations.

The general phenomenon of nonlinear electro-osmotic flow around a
polarizable surface has been termed ``induced-charge
electro-osmosis'' (ICEO)~\cite{iceo}. It arises in a variety of
situations involving polarizable surfaces,  producing circulating
flow patterns similar to those in our problem of a fluctuating
driven membrane. What we now call ICEO flow was first described by
V. Murtsovkin and collaborators in Russia~\cite{murtsovkin} in the
case of metallic colloidal spheres. Recent interest in the subject
has focused on novel phenomena in microfluidic devices, such as AC
electro-osmotic flow around electrode arrays~\cite{ramos,ajdari},
ICEO flow around metal posts~\cite{levitan,harnett} and dielectric
corners~\cite{thamida,yossifon} and induced-charge
electrophoresis~\cite{iceo,gangwal}. However, we are not aware of
any prior theory or experiment describing ICEO around membranes.

The classical theory of electrokinetic phenomena assumes a
constant surface charge, or equivalently, a constant voltage (zeta
potential) between the shear plane at the surface and the
quasi-neutral bulk electrolyte just outside the diffuse charge
layer~\cite{EK}. In that case, the presence of a tangential
component of the electric field
%
(approximately constant across the thickness of the double layer)
leads to electro-osmotic flow that
is linear in the field. For thin double layers, the effective
hydrodynamic slip outside the double layer is given by the
Helmholtz-Smoluchowski formula \cite{EK}
\begin{equation}\label{Helmholz-Smoluchowski}
\vp^*=-\frac{\epsilon \zeta^*}{\eta^*} \Ep^*,
\end{equation}
where $\zeta^*$ denotes the zeta potential across the diffuse
part. This result holds in the asymptotic limit of thin double
layers. It is also valid even if a normal current drives the
diffuse charge out of equilibrium -- all that is required is for
the viscosity and permittivity to be constant within the double
layer and for the bulk salt concentration to be uniform (without
tangential gradients)~\cite{iceo,zaltzman}. At a polarizable
surface, the zeta potential and tangential field component vary in
response to perturbations of the system. This results in nonlinear
ICEO flows which typically vary with the square of the applied
voltage.

\begin{figure*}
\includegraphics[height=2.1in]{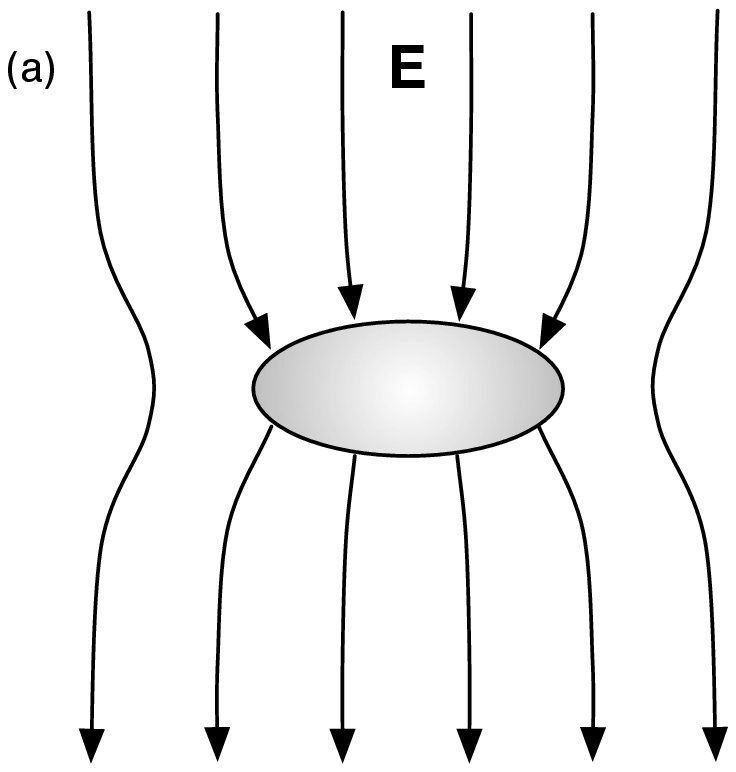} \hspace{0.1in}  \nolinebreak
\includegraphics[height=2.1in]{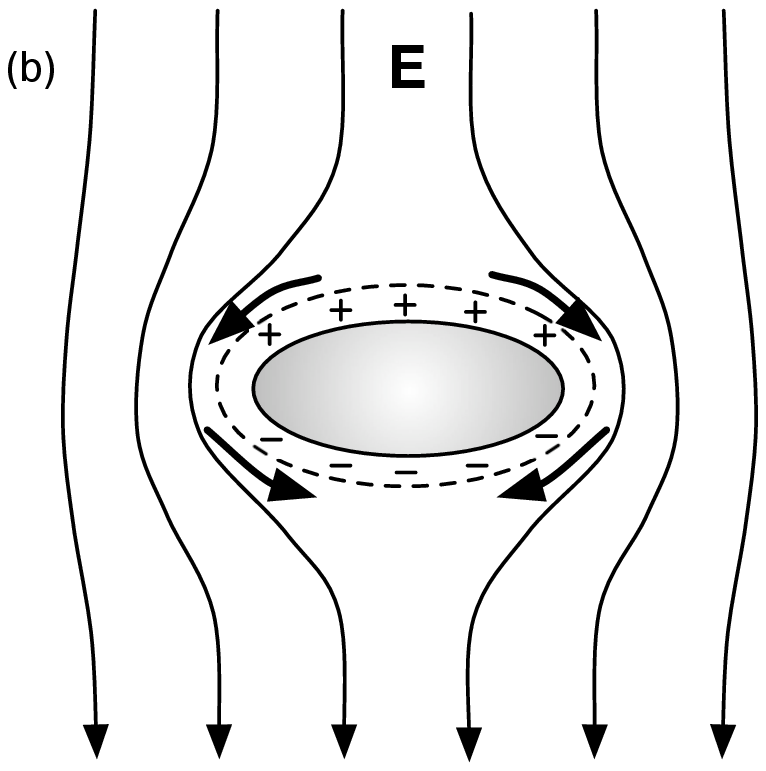}\hspace{0.1in} \ \nolinebreak
\includegraphics[height=2.1in]{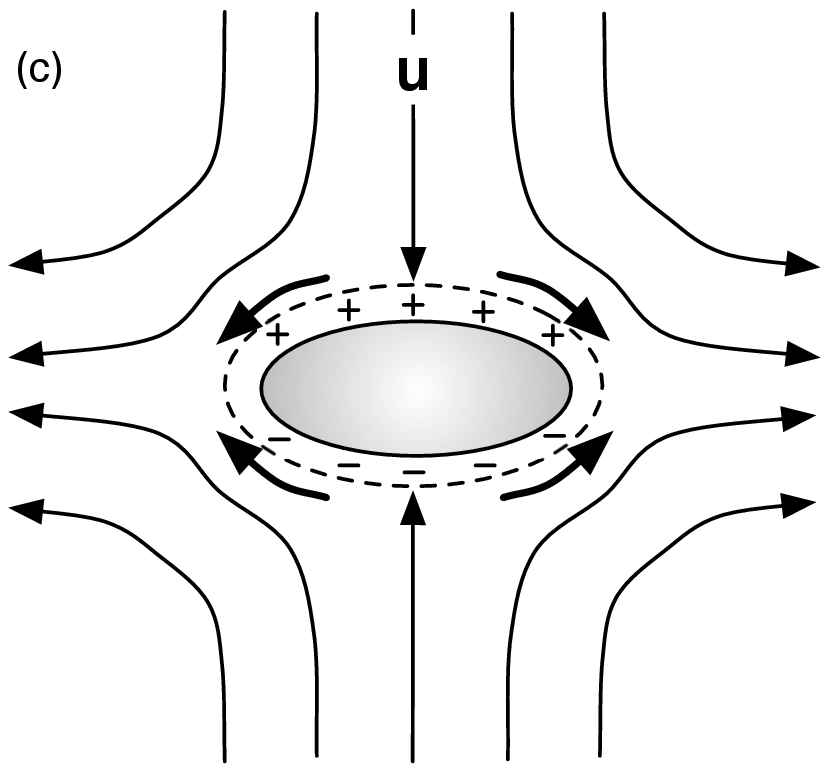}\\
\ \\
\includegraphics[height=2.1in]{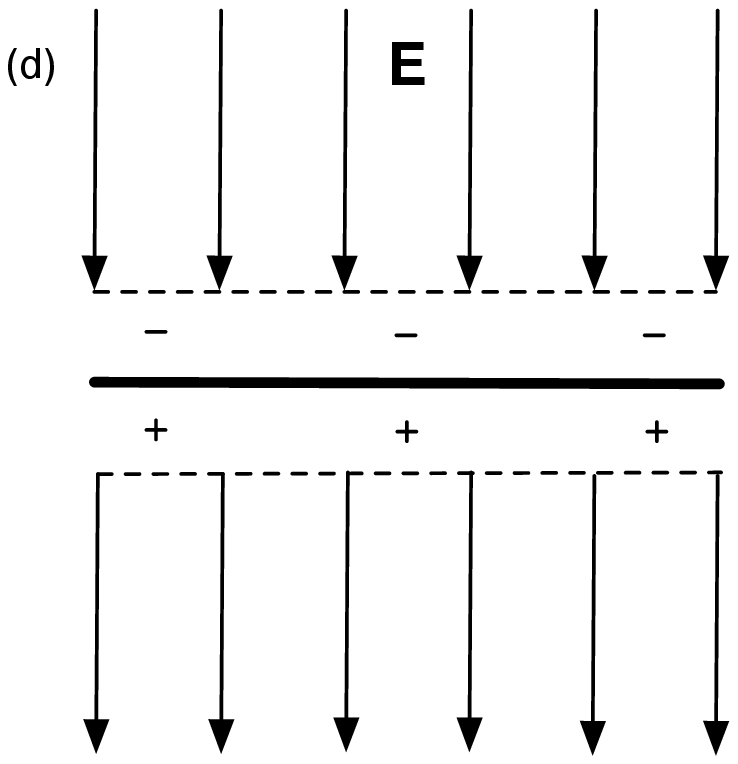} \hspace{0.1in}  \nolinebreak
\includegraphics[height=2.1in]{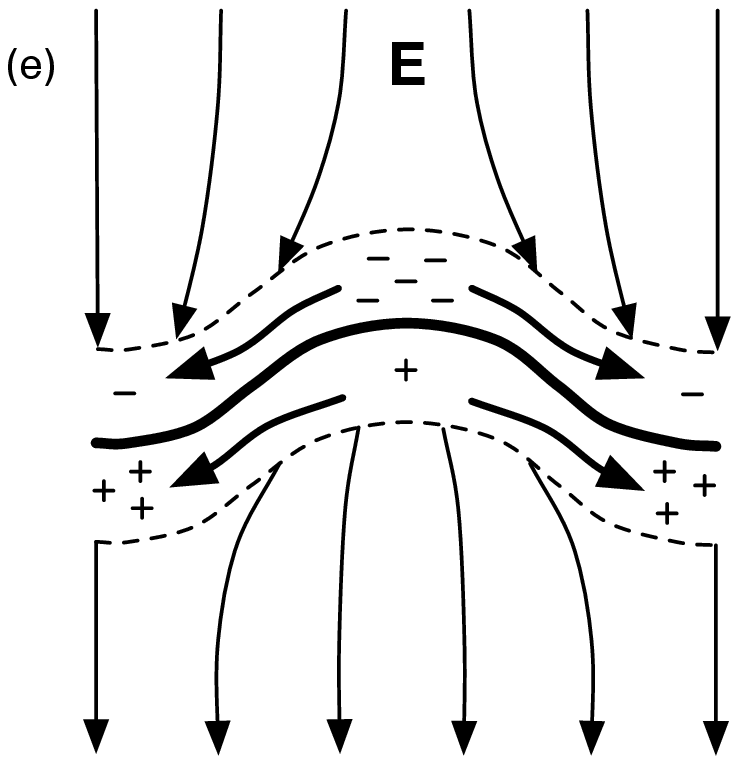}\hspace{0.1in} \ \nolinebreak
\includegraphics[height=2.1in]{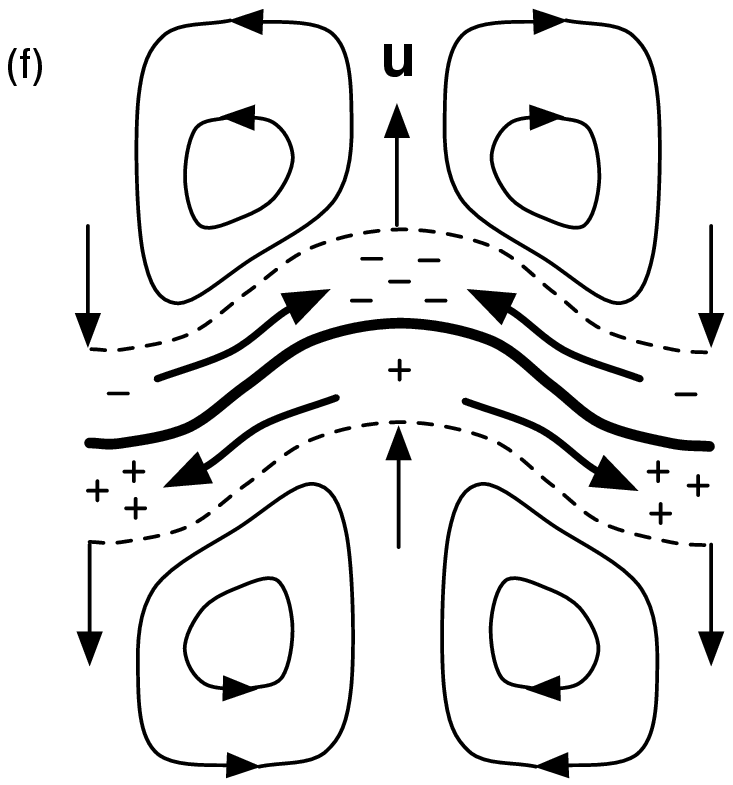}\\
\caption{ The basic physics of ICEO around an ideally polarizable
  metal post ~\cite{iceo} in (a-c), contrasted with our new example a
  driven membrane in (d-f). (a) The metal post is subjected to an
  electric field; (b) capacitive charging of the double layers screens
  the post and thus creates a tangential field (thick arrows); (c) the
  field acts on the induced diffuse charge to produce electro-osmotic
  slip (thick arrows) directed from the peak to the sides. (d) The
  membrane is subjected to an electric field, which drives a current
  through it, and creates small diffuse charge of the opposite sign
  (in the case of the first model of this paper, for which the membrane
  has a zero thickness); (e) the membrane shape fluctuates,
  inducing a shift in the diffuse charge and a tangential field (thick
  arrows); (f) the induced field acts on the initial charge to drive
  ICEO flow (thick arrows) in the reverse direction, from the valleys
  to the peaks. }
\label{fig:iceo}
\end{figure*}

To understand the appearance of ICEO flow in our system, we begin
by considering dominant balances in the dimensionless equations.
We first consider the limit of thin double layers compared to the
perturbation wavelength, $\kps \ll 1$ (or $\kps^* \ll \kappa$),
which is relevant for biological membranes.
In our system of
normalized units, the electro-osmotic slip formula
(\ref{Helmholz-Smoluchowski}) predicts the scaling
\begin{equation}\label{Helmholz-Smoluchowski-normalized}
\vp(\kp,z \geq 1)\simeq -\zeta \Ep (\kp,z \to 0^+),
\end{equation}
where the effective slip velocity outside the double layer ($z \geq 1$) is
proportional to the typical tangential electric field in the diffuse layer, set by its typical value within the Debye layer at the surface ($z \to 0^+$). Although the scaling is the same, a subtle difference with Helmholz-Smoluchowski theory is that the tangential electric field is confined to the diffuse layer and vanishes in the neutral bulk electrolyte ($z \geq 1$).

The expression (\ref{Helmholz-Smoluchowski-normalized}) can be verified by direct integration
of the Stokes equation as follows. After projecting
Eq.~\ref{Stokes} in the transverse direction, and retaining only terms of first order,
one obtains
\bb \triangle \vp + Q(z) \Ep^{(1)}  =  0, \en
 which can be simplified using the condition $\kps \ll 1$ to give
\bb \partial_z^2 \vp (\kp,z) + Q(z) \Ep^{(1)}(\kp,z) =0. \label{vp1} \en
Now from (\ref{BC
potential z>0}) in the limit $\kps \ll 1$,
\bb \label{ep1}
\Ep^{(1)}(\kp,z) = -i \kp \psi^{(1)}(\kp,z)
= -i \kp \sigma h(\kp) e^{-z}.
\en
After inserting Eq~\ref{ep1} into Eq.~\ref{vp1}, using the no-slip
condition $\vp^{(1)}(\kp,z=0^+)=0$, the transverse first-order
velocity profile is found to scale as (dropping a numerical prefactor of $i/4$)
\bb \label{approx vperp}
\vp^{(1)}(\kp,z) \simeq \sigma^2 \kp h(\kp) (1-e^{-2z}),
\en
a scaling which is also confirmed by our solution of the Stokes equation
given in the previous section (cf. Appendix B).

Thus, the scaling of the Helmholz-Smoluchowski relation (Eq.~\ref{Helmholz-Smoluchowski-normalized})
indeed holds with $\zeta=\zeta^{(0)}=-Q^{(0)}(z=0^+)=\sigma$. The only difference is the dropped factor of $1/4$, which results from the decay of the tangential electric field within the diffuse layer, in contrast to the Helmholz-Smoluchowski assumption of a uniform field applied in the bulk electrolyte. Note also that  the first-order perturbed field due to membrane displacement acts on
the leading-order base-state diffuse charge to drive
electro-osmotic flow. This thus differs from other examples of
ICEO flow~\cite{iceo}, where the field acts on the perturbed
charge, as discussed below. Taking into account the constant low-voltage capacitance of the
diffuse layer $C_D^*=\epsilon \kappa$, the induced zeta potential is
related to the total diffuse charge by $\zeta^*=\sigma^*/C_D^*$
\cite{ajdari,bazant2004}. In dimensionless units, we recover again
$\zeta^{(0)} \simeq \sigma$.
Using the
incompressibility condition ~(\ref{incompressibility}),
we obtain the scaling
$v_z(\kp,z \to 0^+) \simeq \kps^2 h(\kp)
\sigma^2$. As illustrated in Fig.~\ref{fig:thinDLflow}(b), the
normal velocity is smaller than the tangential velocity by a
factor $\kps$. Applying the boundary conditions $v_z(\kp,z
\rightarrow 0^+) =
\partial h/\partial t$ together with equations ~(\ref{mb delta free
  energy}-\ref{eq motion 2}), we find that the velocity estimated from
this ICEO argument indeed corresponds to $\Gamma \simeq \sigma^2$
in the effective free energy of the membrane. Note that the
velocity $v_z(\kp,z \rightarrow 0^+)$ scales with the square of
the applied electric field, so ICEO is relevant for both DC and AC
electric fields, as long as the AC period exceeds the charging
time (see below).

An equilibrium term in $\kps^3$, originating in unscreened
dipole-dipole interactions, is also obtained in the calculation of
Ref.~\cite{lomholt2}, but in the high $\kps$ limit, in which $\kps
L \gg 1$. It is absent in the low $\kps$ limit. Since the term we
derive is obtained after taking $L \rightarrow \infty$, it is
clear that the origin of this term is very different in both
calculations and has an explicitly non-equilibrium origin in our
approach.

In the remainder of this section, we give simple scaling arguments
(with dimensions, for clarity) to highlight the basic physics of
this new phenomenon of ICEO that we predict around driven
membranes. For comparison, we first review the canonical example
of ICEO flow around an ideally polarizable, uncharged metal post
in a suddenly applied DC field $E^*$ ~\cite{iceo}, illustrated in
Fig.~\ref{fig:iceo}(a-c). We scale the geometry of the metal post
to that of our curved membrane with an extent $h^*$ parallel to
the field and $\kps^{*-1}$ perpendicular to the field. In the base
state (a) at $t=0$, the metal post is an equipotential surface,
but this is not a stable situation, since the surface is assumed
not to pass any current. Instead, the normal current entering the
diffuse layer charges it locally like a capacitor, until all the
field lines are expelled (after the ``RC'' charging time $\tau^*_c
\sim (D\kappa \kps^*)^{-1}$, where $D$ is a characteristic ionic
diffusivity ~\cite{bazant2004}).

The induced voltage across the
diffuse layer scales as the background voltage applied across the
post, $\zeta^* \sim E^* h^*$. The induced tangential electric
field wraps around the post as shown in (b) and scales as
$\Ep^{*}\sim E^* h^* \kp^*$. Substituting into the slip formula
(\ref{Helmholz-Smoluchowski}) then yields the scaling of the ICEO
velocity
\begin{equation}
\vp^*_{metal} \sim \frac{\epsilon \kp^* h^{*2}}{\eta^*} E^{*2},
\end{equation}
which flows in along the field axis toward the peak of the post
and outward along its surface, as shown in (c).

In our model membrane, the ICEO flow is different in several
important ways, although it shares the same basic principle of an
applied field acting on its own induced diffuse charge around a
polarizable surface. The physical picture is sketched in
Figure~\ref{fig:iceo}(d-f). In this paper, we ignore
diffuse-charge dynamics and focus on the steady response to shape
perturbations. Initially, a normal field $E^*=E_z^{(0)*}$ is
applied to the flat membrane to pass a current through it, as
shown in (d). This induces a zeta potential scaling as
$\zeta^{(0)*} \sim -E^* \kappa^{-1}$ of opposite sign to the
ideally polarizable metal post, due to the much lower ``inner''
capacitance of the membrane compared to the ``outer'' capacitance
diffuse layers ($\delta_m \gg 1$), as explained above.

Now consider a fluctuation in the shape of the membrane, as shown
in (e). Since $\delta_m\gg 1$, the membrane carries most of the
voltage applied to the total double layer, so the perturbation of
the induced zeta potential scales as $\zeta^{(1)*} \sim - E^* h^*$
since there is a transfer of this voltage (or the corresponding
diffuse charge $q^{(1)*}\sim - \epsilon \kappa \zeta^{(1)*} =
\epsilon E^* \kappa h^*$) from the diffuse layer on the protruding
side to that of the other side.

As shown in (e), the induced
tangential field, scaling as $\Ep^{(1)*} \sim \zeta^{(1)*} \kp^*$,
is the same on both sides of the membrane (even in $z$) and
directed from the peaks ($h>0$) to the valleys ($h<0$) of the
shape fluctuation. It may seem surprising that the field is bent
away from the extra negative induced charge in the diffuse layer
near the peak and toward the extra positive induced charge in the
diffuse layer in the valley, but this is due to the large bound
positive (negative) charge on the upper (lower) side of the
membrane, which greatly exceeds the diffuse charge in the regime
$\delta_m \gg 1$). Ignoring the small diffuse charge, it becomes
clear that the field is mainly perturbed to avoid the protrusion
of the positively charged membrane.

Substituting these estimates in (\ref{Helmholz-Smoluchowski}), we
obtain the basic scaling of the ICEO velocity
\begin{equation}
\vp^*_{membrane} \sim -\frac{\epsilon \kp^* h^{*}}{\eta^* \kappa}
E^{*2} = -\frac{\vp^*_{metal}}{\kappa h^*}    \label{eq:vmem}
\end{equation}
As in the example of the metal post, the ICEO flow around the
membrane increases with the aspect ratio of the shape
perturbation, $\kps^* h^*$, since it is associated with
protrusions in the field direction. Compared to the ideally
polarizable metal post (c), however, the curved membrane (f)
exhibits ``reverse'' ICEO flow, which is directed from the valleys
to the peaks. It is also reduced by a factor $\kappa h^*$, which
shows that ICEO flow around a driven membrane is inherently a
phenomenon of thick double layers (compared to the shape
perturbation amplitude). Although these flows are weak compared to
large-scale ICEO flows in microfluidics and colloids in similar
geometries, we have seen that they are strong enough to make a
significant contribution to the small-scale dynamics of
fluctuating biological membranes.

\begin{figure}
{\par { \rotatebox{0}{\includegraphics[scale=0.9]{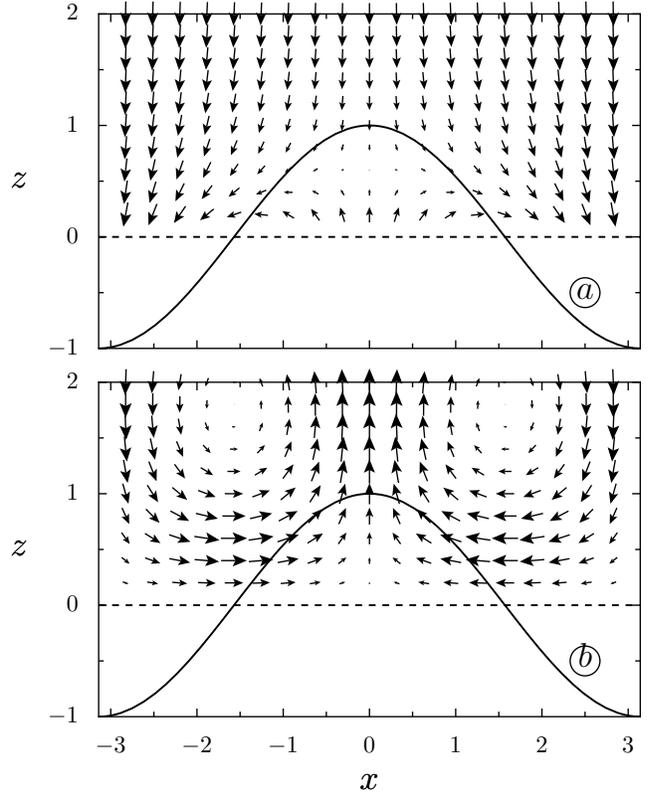}} } \par}
  \caption{ Induced-charge electro-osmotic (ICEO) flow around a driven
    membrane in the regime of thick double layers $\kps = \kps^* /
    \kappa^* \approx 1$, where the thickness of the Debye layer is
     comparable to the wavelength of the shape
    perturbation and to the amplitude of modulation of the membrane.
    Indeed in units of the Debye layer, the amplitude of modulation
    of the membrane is 1, the wavevector is also 1
    and the applied field corresponds to $\sigma=8$.
    (a) The total electric field ${\bf E}$ avoids the
    protrusion peak due to the large positive bound charge on the
    membrane, even though the (much smaller) induced charge in the
    diffuse layer is negative at the peak and positive in the valley.
    (b) Vortices of induced-charge electro-osmotic flow, scaling as
    $E^2$, and driven from the valleys to the peaks by tangential
    fluid forces as in the thin double layer case of
    Fig.~\ref{fig:thinDLflow}, but with stronger effects of normal
    forces in the recirculating regions, as explained in the text.}
\label{fig:thickDLflow}
\end{figure}

The physical mechanism sketched in Fig.~\ref{fig:iceo}(d-f) can be
seen more clearly in Fig.~\ref{fig:thickDLflow}, where the
electric field and ICEO flow are shown for a shape perturbation of
higher curvature with $\kps^* = h^*= \kappa$, where the
double-layer thickness is comparable to the perturbation
wavelength. In this regime, ICEO flow can no longer be understood
purely as an effective slip given by
(\ref{Helmholz-Smoluchowski}), since normal forces on the fluid in
Eq.~(\ref{force components}) also play an important role in the
flow. As described above, normal forces contribute to membrane
motion and thus viscous flow of the type in
Fig.~\ref{fig:thinDLflow}(a), but they also produce osmotic
pressure, which can drive flow relative to the instantaneous
membrane position. For a thin quasi-equilibrium double layer,
tangential gradients in osmotic pressure are balanced by
electrical forces within the double layer and do not contribute to
effective slip, as long as the bulk salt concentration is
uniform~\cite{iceo,zaltzman,EK}. For thick double layers, however,
normal forces can also contribute to the flow, mainly within a
distance of $\kappa^{-1}$ from the peaks and valleys, and the
associated flows have same scaling as Eq.~ (\ref{eq:vmem}).

A detailed study of how ICEO around a driven membrane depends on all
the dimensionless parameters in our model would be interesting,
but here we have focused on the regime $\kps \ll 1$, $\delta_m \gg 1$, $d \ll 1$ and $\delta_m \gg 1$. This regime corresponds to the first model discussed in this paper of a membrane
of zero thickness and zero dielectric constant.
As discussed in section \ref{zero thickness}, the sign of the charge distribution
also depends on how conductive the membrane is as compared to the electrolyte. In real
biological membranes the conditions $\delta_m \gg 1$ and $\delta_m \ll 1/G$ both hold simultaneously.
In that case, due to the latter condition $\delta_m \ll 1/G$, the sign of
zeroth order charge distribution is reversed as compared to that obtained in the first model of the paper, as shown in Fig.~\ref{fig:potential zero thickness} and Fig.~\ref{fig:iceo}d.
To summarize, to adapt these figures to the more biologically relevant case, one should reverse
the sign of the charge distribution and that of the first order correction to the potential. Fortunately,  since the ICEO flow velocity scales as the square of the electric field, the direction of the fluid flow shown in Fig.~\ref{fig:iceo}f will always be correct irrespective of the sign of the diffuse charge distribution.

\section{Effects due to inhomogeneities in the pumps/channels concentration}
\label{sec: concentration fluct}

We now discuss the effect of
including the spatial dependance of the concentration field of the
channels or pumps. The membrane free energy is modified by this
concentration field. It is now written as \bb \label{mb free
energy} F_{mb}=\frac{1}{2} \int d^2 \rp [ \kappa_0 \left( \nabla^2
h \right)^2 + \sigma_0 \left( \nabla h \right)^2 - 2 \Lambda \phi
\nabla^2 h + \beta \phi^2 ]. \en The new parameters $\Lambda$ and
$\beta$ are the curvature-coupling coefficient and the
compressibility associated with the channel concentration field
respectively. Note that $\phi$ represents the deviation of the
concentration field with respect to the uniform concentration. For
simplicity, terms such as $(\nabla \phi)^2$, describing the energy
cost of a non-uniform concentration field, have been ignored.

The equation of motion for the $\phi$ field follows from the above
form of the membrane free energy. Gradients of chemical potential
defined as $\mu_{mb}=\partial F_{mb}/
\partial \phi$ provide the driving force for the motion of the
channels on the membrane. Thus \cite{misbah}, \bb
\frac{\partial \phi(\kp)}{\partial t}=- \kps^2 [ \Lambda \kps^2
h(\kp) + \beta \phi(\kp) ] \label{eq of motion for phi}. \en

In a linear approximation, the electrostatic potential can be
written as a superposition of a base state contribution
$\psi^{(0,0)}$, a contribution linear in the membrane height field
$\psi^{(1,0)}$ and a contribution linear in the channel
concentration field $\psi^{(0,1)}$. Augmenting our previous
notation, $q^{(0,0)}=Q$, $c_k^{(0,0)}=N_k$ and $\psi^{(0,0)}=\Psi$
in the base state, we now have, instead of equation~(\ref{def potential})
\begin{equation}\label{def potential double perturbation}
\begin{array}{ccc}
\psi(\kp,z) = \Psi(z) + \psi^{(0,1)}(\kp,z) + \psi^{(1,0)}(\kp,z), \\
q(\kp,z) = Q(z) + q^{(0,1)}(\kp,z) + q^{(1,0)}(\kp,z), \\
c_1(\kp,z) = N_1(z) + c_1^{(0,1)}(\kp,z) + c_1^{(1,0)}(\kp,z), \\
c_2(\kp,z) = N_2(z) + c_2^{(0,1)}(\kp,z) + c_2^{(1,0)}(\kp,z).
\end{array}
\end{equation}
The equations obeyed by $\psi^{(1,0)}$ and $q^{(1,0)}$, as well as
$\psi^{(0,1)}$ and $q^{(0,1)}$ follow from equation~(\ref{eq 1st order
charge fluctuation}).

For the contribution linear in the concentration field
$\psi^{(0,1)}$ of the channels, the boundary conditions at the
membrane surface impose continuity of the ion fluxes in the
channels. Using the linear form for the conductances, we have \bb
J_k= G_k \Delta \mu_k = \left( G_k^{(0,0)} + G_k^{(0,1)} \phi
\right) \left( \Delta \mu_k^{(0,0)} + \Delta \mu_k^{(0,1)} \phi
\right). \en Collecting terms linear in $\phi$, we obtain \bb
J_k^{(0,1)}=\alpha_k^{(0,1)} \phi \en with \bb
\alpha_k^{(0,1)}=G_k^{(0,1)} \Delta \mu_k^{(0,0)} + G_k^{(0,0)}
\Delta \mu_k^{(0,1)}. \en We assume $\alpha_k=\alpha$, which
represents the pumping rate. This condition only needs to be
enforced at the unperturbed interface and along the $z$ direction
so that \bb \left(
\partial_z q^{(0,1)} +
\partial_z \psi^{(0,1)} \right)_{z=0}  =
 \alpha \phi(\rp). \label{BC flux simple 01} \en

We find the following solution for the electrostatic potential
with these boundary conditions: for $z>0$, \bb
\psi^{(0,1)}(\kp,z)=-q^{(0,1)}(\kp,z)= \alpha \phi(\kp) \left(
\frac{e^{-lz}}{l} - \frac{e^{-k_\perp z}}{k_\perp} \right), \en
and for $z<0$, \bb \psi^{(0,1)}(\kp,z)=-q^{(0,1)}(\kp,z)= \alpha
\phi(\kp) \left( \frac{-e^{lz}}{l} + \frac{e^{k_\perp z}}{k_\perp}
\right). \en Note that the corrections to the charge distribution
and potential are odd functions of $z$ in this case. This can be
understood from the fact that the ion channels locally create a
depletion of ions on one side and an increase of ion concentration
on the other side. This depletion can be quantified through the jump in
concentration of the charges across the membrane \bb
q^{(0,1)}(\kp,0^+)-q^{(0,1)}(\kp,0^-)=-\frac{2 \alpha
\phi(\kp)}{l}. \en This jump in concentration provides an osmotic
pressure difference between the two sides of the membrane, whose
effect is irrelevant, however, since we
have assumed the absence of permeation in writing the boundary
condition of equation~(\ref{permeation}) \cite{PRE_JBManneville}.

The concentration field $\phi$ enters the equation of motion of
the height field only through the membrane restoring force
$\partial
 F_{mb}/ \partial h$. This is because terms proportional to $\phi$
cancel in the difference of the stress along the $z$ direction
between both sides of the membrane in equation~(\ref{BC for 1st order}),
due to the fact that $\psi^{(0,1)}(\kp,z)$ is an odd function of
$z$. As a consequence of this simplification, the transport coefficient
$\alpha$ introduced in equation~(\ref{BC flux simple 01}) does not enter
the equation of motion for $\phi$ or for $h$. With equation~(\ref{eq
of motion for phi}), and the equation of motion for $h$,
\begin{eqnarray}
 \frac{\partial h(\kp)}{\partial t} & = &
-\frac{1}{4 \kps } [ \left( (\sigma_0+\Sigma)h(\kp) + \Lambda
\phi(\kp) \right) \kps^2 \\
& + & h(\kp) \left( \kappa_0 + K \right) \kps^4 + h(\kp) \Gamma
\kps^3 ], \label{eq of motion for h_2}
\end{eqnarray}
the condition of stability of the membrane with its inclusions may
be obtained, provided the bare elastic moduli of the membrane, and
the induced surface charge $\sigma$, are known.

\section{Electrically driven membrane of finite thickness}
\label{finite thickness} In this section, we consider a bilayer of
finite thickness $d$ and dielectric constant
$\epsilon_m<\epsilon$. There is then an electrical coupling
between the membrane and the surrounding electrolyte, with a
strength measured by the parameter $t=\epsilon_m/(\kappa d^*
\epsilon)=\delta_m^{-1}$~\cite{helfrich,andelman}. For equilibrium
membranes, the importance of this coupling is discussed in
refs.~\cite{kleinert,chou}. In dimensionless units, this coupling
becomes $r/d$, where $r=\epsilon_m/\epsilon$ and $d$ is the
dimensionless membrane thickness.

For non-equilibrium driven membranes, capacitive
effects associated with the finite thickness of the
membrane dominate electrostatic corrections to the
membrane elastic moduli, except at low ionic strength
\cite{lacoste}. Further, capacitive effects are essential
to explain voltage induced motion in cell membranes
containing ion channels \cite{sachs} and shape transitions
of giant vesicles in AC electric fields \cite{lipowsky}.
We ignore variations in the concentration of the channels
here~\cite{lacoste}. We will also only consider the mode of fluctuation of
the membrane in which each layer of the membrane fluctuates
in phase with respect to each other, so that the position
of each layer is $\pm d/2 + h(\rp)$.

For simplicity, we only discuss the case of symmetric
electrolytes: $n^-=n^+$, $D_1=D_2=D$, and $G_1=G_2=G$. We denote
by $\psi_m$ the internal potential for $|z|<d/2$ and $\psi$ the
electrolyte potential for $|z|>d/2$. When $t\neq0$, the boundary
conditions at the membrane are modified, becoming
\begin{equation}\label{BC1}
\begin{array}{c}
  \partial_z \psi^{(0)}(z \to \pm d/2)=r \partial_z
\psi_m^{(0)}(z \to \pm d/2), \\
  \psi^{(0)}(z \to \pm d/2)=\psi_m^{(0)}(z \to \pm d/2).
\end{array}
\end{equation}
The first equation is the continuity condition for the normal
electric displacement and the second equation is the continuity
condition of the potential.

We solve the analog of
equations~(\ref{charge cons lin}-\ref{charge cons lin2}) together with an
additional equation describing the region in between the bilayer.
In this intermediate region, it is assumed that there is no charge
density.

We find the following solution for the base state
\begin{equation}\label{Efield-finite d}
\begin{array}{c}
  E_z^{(0)}(z)=-\sigma  -\tilde{\sigma}  \exp{
  \left( z+d/2 \right) }, \,\,\, {\rm for } \,\,\, z<-d/2 \\
  E_z^{(0)}(z)=-\sigma  -\tilde{\sigma}  \exp{
  \left( -z+d/2 \right) }, \,\,\, {\rm for } \,\,\, z>d/2 \\
E_m^{(0)}=-\frac{\sigma + \tilde{\sigma}}{r}, \,\,\, {\rm for }
\,\,\, -d/2<z<d/2.
\end{array}
\end{equation} Here $\sigma$ still represents the surface charge of the Debye layers
which is defined as in the case of zero thickness
in equation~(\ref{def_sigma}). The current versus voltage relation
obtained for zero thickness in equation~(\ref{final currents}) still
holds in the finite thickness case, once $L$ is replaced by
$L-d$. Similarly equation~(\ref{exp current}) holds after replacing $L^*$
by $L^*-d^*$, while equation~(\ref{sigma_with dimension}) holds
unchanged. In equation~(\ref{Efield-finite d}), we have introduced a new
quantity $\tilde{\sigma}$ with the following property \bb
\tilde{\sigma}=\int_{d/2}^\infty Q^+(z) dz=-\int_{-\infty}^{-d/2}
Q^-(z) dz. \en Note that $\sigma$ and $\tilde{\sigma}$ are related
to each other by \bb \label{rel sigmas} \tilde{\sigma}=\frac{r
\left(\sigma d - \sigma L +V \right) - \sigma d}{2 r + d}. \en
When  dimensions are reinstated, one can see that only the
diffusion time of the ions within a Debye layer enters in
$\sigma^*$ whereas $\tilde{\sigma}^*$ also contains the RC
characteristic time of the membrane \cite{lacoste}.

The boundary conditions for the first order correction in the
membrane height field to the electrostatic potential and to the
charge density are
\begin{eqnarray}\label{BC finite d1}
\partial_z \psi^{(1)}(\kp,\pm d/2) & = & \mp h(\kp)
 \partial_z^2 \psi^{(0)} (d/2) \nonumber \\
& + &    r \partial_z \psi_m^{(1)} (\kp,\pm d/2),
\end{eqnarray}
\begin{eqnarray}\label{BC finite d2}
\psi^{(1)}(\kp,\pm d/2) & = & \psi_m^{(1)}(\kp,\pm d/2) \nonumber \\
& + &  h(\kp) \left(\partial_z  \psi_m^{(0)} - \partial_z
\psi^{(0)} \right) (d/2),
\end{eqnarray}
\bb \psi^{(1)}(\kp,\pm \infty)=q^{(1)}(\kp,\pm \infty)=0, \en \bb
\partial_z q^{(1)}(\kp,\pm d/2)  + \partial_z \psi^{(1)}(\kp,\pm d/2)  = 0. \en

The last equation corresponds to the boundary condition
of zero flux in the first order solution. This condition
was used
in equation~(\ref{BC flux simple 10}). The equations to first order in
the height in the electrolyte regions $|z|>d/2$ are
\begin{eqnarray}\label{eq 1st order charge fluctuation finite d}
\left( \partial_z^2 -k_\perp^2 \right) \psi^{(1)}(\kp,z) +q^{(1)}(\kp,z) & = & 0, \\
\left( \partial_z^2 -k_\perp^2 \right) \left(q^{(1)}(\kp,z) +
\psi^{(1)}(\kp,z) \right) & = & 0,
\end{eqnarray}
subject to the boundary conditions given above.

The equations in the inside medium for $z<|d/2|$ are
$q^{(1)}(\kp,z)=0$ and \bb
\left(
\partial_z^2 -k_\perp^2 \right) \psi_m^{(1)}(\kp,z)=0. \en
For $z>d/2$ and $z<-d/2$, $q^{(1)}(\kp,z)$ and $\psi^{(1)}(\kp,z)$
remain of the form $A \exp{ \left( \mp z + d/2 \right) }$, where
$A$ is a complicated function of $\kps, r, d, \sigma$ and
$\tilde{\sigma}$. The first order correction to the potential in
the inside medium has the form \bb \psi_m^{(1)}
(\kp,z)=\psi_m^{(1)} (\kp,d/2) \frac{e^{\kps d/2} \left( e^{\kps
z}+ e^{- \kps z} \right)}{e^{\kps d}+1}. \en

\begin{figure}[htbp]
  \centering
\rotatebox{0}{\includegraphics[scale=0.7]{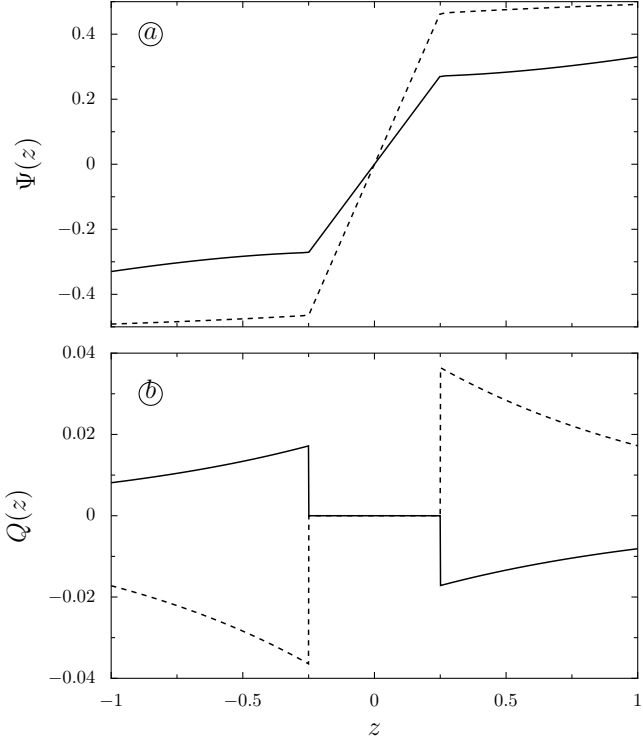}}
\caption{Solutions of the electrokinetic equations for a
    membrane of finite thickness and symmetric ion concentrations.
The electrostatic potential $\Psi(z)$ is shown in (a) and the
quantity $Q(z)$ (which represents half the charge distribution) is
shown in (b). We use dimensionless units and the following parameter values
$L=100$, $V=2$, $G_1=G_2=1$, $d=0.5$, $r=1/40$ and $\kps=1$ for
the solid line, and the same parameters except for $G_1=G_2=0.01$
for the dashed line. For convenience, the potential represented in the
solid line of (a) has been multiplied by an arbitrary factor of 10.
For the solid line in (a), the curvature of the potential is positive near $z=d/2$,
as in figure \ref{fig:potential zero thickness}a, this corresponds to a negative value of
$\tilde{\sigma}$. The charge distribution is as shown in the solid line of (b) and similar to figure \ref{fig:potential zero thickness}b.
For the dashed line in (a), the curvature of the potential is negative near $z=d/2$,
this corresponds to a positive value of $\tilde{\sigma}$. Note that the sign of the charge distribution (dashed line in (b)) is reversed as compared to the solid line in (b).}
 \label{fig:potential finite thickness}
\end{figure}

In Figure~\ref{fig:potential finite thickness}, the potential
$\Psi(z)$ and the quantity $Q(z)$ (which represents half the charge distribution) are shown in dimensionless units for two choices of
parameters. These parameters correspond to a positive and a
negative value of $\tilde{\sigma}$. The potential profiles
illustrated in Figure~\ref{fig:potential finite thickness} show
that the sign of the electric field at the membrane surface ($\sim
-\nabla \psi$) is controlled by the sign of $\tilde{\sigma}$. As
shown in the appendix, in this case of a symmetric electrolyte and
with the approximation $G_1=G_2=G$ the sign of $\tilde{\sigma}$ is
positive when $\delta_m < 1/G$ and negative otherwise. The
condition $\delta_m < 1/G$ is equivalent to $\epsilon_m/\epsilon
\gg G^*/G_0^*$, where $G^*$ is the typical conductance of typical
ion channels/pumps and $G_0^*$ is the conductance of a layer of
electrolyte of thickness $d^*$. For real biological membranes, the membrane is
typically much less conductive than the surrounding medium and thus
$\delta_m < 1/G$. So the charge distribution in this case should be as the
dashed line of Figure~\ref{fig:potential finite thickness}b, which has we mentioned earlier, has the opposite sign as compared to the charge distribution shown in Fig.~\ref{fig:potential zero thickness}.

\subsection{Electrostatic corrections to elastic moduli}
The Stokes equations can be solved as before, although the
computations are now more involved. With the potential
at zeroth and first order computed above, we first
construct $\vf (\kp, z)$ using equations~(\ref{force
components}). After Fourier transforming in $z$, we insert
the result in equation~(\ref{vz}-\ref{pressure}). From the
solution of the Stokes equations, the velocity is obtained
everywhere in the domain $|z|>d/2$.

We now have two boundary conditions for the stress
tensor $\tau_{zz}^{(1)}$ at $z=\pm d/2$. Solving
for the velocity field and extrapolating this
velocity to $z=0$, an effective free energy of
the same form as in equation~(\ref{mb delta free
energy}) is obtained. The tension obtained from this
calculation is the same as the one calculated from
equation~(\ref{surfacetension}-\ref{surfacetensionb}).

Both methods yield the result that the surface tension is the
sum of an internal contribution $\Sigma_{in}$, arising from the
contribution of the field lines which penetrate within the membrane,
and an external contribution $\Sigma_{out}$ associated to the field
lines present in the Debye layers. These take the form\cite{lacoste}
\begin{eqnarray}
\Sigma_{in} &=& -r d E_m^2=-d \frac{\left(\sigma + \tilde{\sigma}\right)^2}{r}, \label{sigma_in} \\
\Sigma_{out} &=& -\tilde{\sigma}^2-4 \sigma \tilde{\sigma} +
 d \sigma^2,
\end{eqnarray}
or \bb \Sigma_{in}^*=-\frac{(\sigma^* + \tilde{\sigma}^*)^2}{t
\kappa \epsilon}, \label{sigma in*} \en and \bb \Sigma_{out}^*
=\frac{-(\tilde{\sigma}^*)^2-4 \sigma^* \tilde{\sigma}^* +
 \kappa d^* (\sigma^*)^2}{\epsilon \kappa}, \label{sigma_out*}
 \en in terms of dimensionful quantities.

The negative contribution $\Sigma_{in}$ is known as the Lippmann
tension~\cite{sachs}. It is usually larger in absolute value than
$\Sigma_{out}$. From equation~(\ref{surfacetension0}), it follows
that $\Sigma_{in}$ is the electromagnetic energy of the internal
field $E_m$ contained within the space of the membrane. Since
$\Sigma_{in}$ is always negative, as illustrated qualitatively in
Figure~\ref{fig:tension}, the total membrane tension
$\sigma_0+\Sigma_{in}+\Sigma_{out}$ can become negative at some
critical value of the internal field $E_m$, leading to the
instabilities discussed in ref.~\cite{pierre}. Note that such an
instability is present at zero wavelength, unlike the finite
wavelength instability discussed from equation (\ref{growth
rate}). Our calculations also yield the moduli \bb \Gamma =
\frac{\sigma \tilde{\sigma}}{2} \left( 8 + d^2 + 4 d \right), \en
and $K$, which is a complicated quadratic function of $\sigma$ and
$\tilde{\sigma}$.

All these moduli reach simple limiting values when the Debye
length goes to zero ($\kappa \rightarrow\infty$ or equivalently
$n^* \rightarrow\infty$). This limit is best understood in
dimensionful notation. From equations~(\ref{exp current}-\ref{sigma_with
dimension}), one finds that $i_k^*=-G_k^* V^*$ and $\sigma^*=0$.
From equation~(\ref{rel sigmas}), we obtain $\tilde{\sigma}=rV/(2r+d)$.
Using equation~(\ref{sigma dim ful-less}) we have \bb
\tilde{\sigma}^*=\frac{k_B T \kappa  \epsilon r V}{e \left( 2r+d
\right)}=\frac{\kappa \epsilon_m V^*}{2r+ \kappa d^*}, \en which
in the limit $\kappa \rightarrow\infty$ goes to $\epsilon_m
V^*/d^*$.

The values of the moduli in this limit are independent of $G$, a
consequence of the fact that electrical currents ($i_k^*$ is
non-zero) are not accompanied by charge accumulation in the Debye
layers, because $\sigma^*=0$. From equation~(\ref{sigma_out*}), we find
that $\Sigma_{out}^*=0$. Using $t=r/\kappa d^*$ in equation~(\ref{sigma
in*}), we find a non-zero limit for $\Sigma_{in}^*=-\epsilon_m
(V^*)^2/d^*$. Thus the limit for the overall tension in the high
salt limit is $\Sigma_0^*=-(V^*)^2 \epsilon_m /d^*$. This
resembles the energy of a plane capacitor with a voltage drop
$V^*$ and thickness $d^*$, although the system is not strictly
analogous to a capacitor since electric currents are
present ($i_k^*$ is non-zero). For the bending modulus, the
limiting value is \bb K_0^*=\frac{5(V^*)^2 \epsilon_m d^*}{24},
\label{K_0*} \en and for $\Gamma$ the limit is $0$. The variation
of $\Sigma^*/\Sigma_0^*$ and of $K^*/K_0^*$ as a function of the
inverse Debye length $\kappa$ are shown in Figure~\ref{fig:hydro
flows} for the case of zero and non-zero conductance.

Another limit of experimental relevance is that of small
conductance $G \rightarrow 0$. In this limit, $\sigma=0$ since
there is no ion current in the medium as a consequence of
$i_k^*$ being zero.
From equation~(\ref{rel sigmas}), we obtain $\tilde{\sigma}=rV/(2r+d)$
which means $\tilde{\sigma}^*=V^* \epsilon \kappa t /(1+2t)$ in
terms of dimensionful quantities. This shows that the membrane is
a capacitor of surface charge $\tilde{\sigma}^*=-\epsilon_m
E_m^*=V^* \epsilon \kappa t / (1+2t)$, where $E_m^*$ is the
internal field \cite{lacoste}. The equivalent circuit for this
problem is composed of three planar capacitors in series. One of
these is the membrane (of capacitance per unit area $\epsilon_m/d$) while the
other two correspond to the Debye layers on each side (of
capacitance $\epsilon \kappa$ per unit area), yielding a total
capacitance $C^*=\epsilon \kappa t/(2t+1)$.

Using equations~(\ref{sigma in*}-\ref{sigma_out*}), we find
$\Sigma_{in}^*=-\tilde{\sigma}^2/ t \kappa \epsilon$,
$\Sigma_{out}^*=-\tilde{\sigma}^2/ \kappa \epsilon$, and
$\Gamma=0$. The same expressions for the surface tension in the
$G=0$ limit were obtained recently using a different method in
Ref.~\cite{lomholt2}. For comparison, we provide
$K^*$ for this case \begin{eqnarray} K^* & = & \frac{1}{24 r
\epsilon \kappa^3} (\tilde{\sigma}^*)^2 ( 18r+5 (\kappa d^*)^3
\nonumber \\ & + & 24 \kappa d^* r + 15 (\kappa d^*)^2 r ).
\label{K*}
\end{eqnarray}
Note that we recover the result of equation~(\ref{K_0*}) when $\kappa
\rightarrow \infty$.

This expression has some similarities as well as some differences
with the expression for the bending modulus given in
ref~\cite{lomholt2}. This discrepancy is likely to originate in
the very different starting points for both calculations: the
results of Ref. \cite{lomholt2} are obtained from an explicitly
equilibrium approach, whose results remain unchanged if
hydrodynamic effects are incorporated. On the other hand, we begin
with an explicitly non-equilibrium problem incorporating
hydrodynamics from the outset.

It is unclear that it should simply suffice to set $G \rightarrow
0$ in our results to recover results derived for the equilibrium
calculation. However, we note that in any case the predictions for
$K$ of both models are numerically very close and there is exact
agreement for $\Sigma$. We also stress that both models predict
that the electrostatic contribution to the bending modulus should
increase with the salt concentration, whereas the electrostatic
contribution to the surface tension should decrease with the salt
concentration as illustrated in the solid lines of
Figure~\ref{fig:tension-kappa}. Both quantities also reach a well
defined limit in the large salt concentration limit (see similar
figure in ref~\cite{lomholt2}).

\subsection{Numerical estimates, capacitive effects and fluctuation spectra}
In the non-conductive limit (capacitor model), we find that
$\Gamma=0$. For $V^*=50$mV, $L^*=1\mu$m, $G=0$,
$D=10^{-5}$cm$^2/s$, $n^*=16.6$mM, $d^*=5$nm and
$\epsilon_m/\epsilon = 1/40$, we have $\kappa^{-1}=2.38$nm, $t=1.2
\cdot 10^{-2}$, $\delta_m=1/t=84$ and $d=\kappa d^*=2.1$,
$V=eV^*/k_BT=1.95$, $L=\kappa L^*=419.7$. The elastic moduli are
$\Sigma_{in}^* = -8.4 \cdot 10^{-6}$ Jm$^{-2}$,
$\Sigma_{out}^*=-1.0 \cdot 10^{-7}$ Jm$^{-2}$, $\Gamma = 0$ and
$K^* = 0.011 k_B T$. Let us now consider instead the case of a
conductive membrane, with $G^*=10 \Omega^{-1}$/m$^2$, a value
typical for ion channels \cite{hille}. This corresponds to a
density of the potassium channels of 0.5 $\mu$m$^{-2}$ and for
$V^*=50$mV, the electrical current going through the membrane is
about 0.5A/m$^2$, which corresponds to about $3 \cdot 10^6$ ions
going through a patch of 1$\mu$m$^2$. The dimensionless channel
conductance is $G=3.8 \cdot 10^{-7}$. This very small value indicates that
the membrane is significantly much less conductive than the electrolyte, and thus
we are typically always in the regime $\delta_m \ll 1/G$ for biological membranes.
We also find that
the order of magnitude of the tension and bending modulus are
unchanged and a small value of $\Gamma^*=8.2 \cdot 10^{-20}$
Jm$^{-1}$ is found \cite{lacoste}. This indicates that the
capacitor model with $G=0$ is a good starting point for the
calculation of the moduli in this case.

The importance of capacitive effects is confirmed by the observation that
the values of the moduli obtained here are much larger than the
corresponding estimates for the zero thickness case. This
can be understood using an equivalent zero thickness model
discussed in appendix A. Also, by varying the ionic
strength in the case where ion transport is present ($G \neq 0$),
we find that the capacitor model holds at high ionic
strength but becomes invalid at low ionic strength, where ion
transport has a stronger impact on the moduli.

\begin{figure} {\par {
\rotatebox{0}{\includegraphics[scale=0.6]{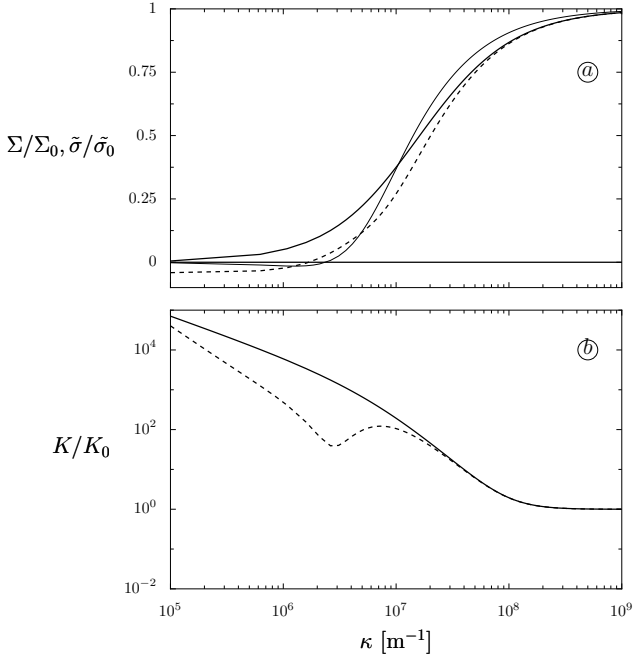}} }
\par}
\caption{Electrostatic contribution to the tension $\Sigma$ and to
the bending modulus $K$ as a function of the inverse Debye length
$\kappa$ in the $G^*=0$ limit (thick solid line) and for $G^*=10
\Omega^{-1}$/m$^2$ (dashed line). (a) Ratios of normalized
electrostatic contribution to the tension $\Sigma/\Sigma_0$
(dashed line for $G^*=10 \Omega^{-1}$/m$^2$ and solid line for
$G^*=0$) and of $\tilde{\sigma}/\tilde{\sigma}_0$ (thin solid line
for $G^*=10 \Omega^{-1}$/m$^2$) are shown as function of $\kappa$.
The tension $\Sigma$ (resp. the surface charge $\tilde{\sigma}$)
are normalized by their value in the infinite $\kappa$ limit
$\Sigma_0$ (resp. $\tilde{\sigma}_0$). Below $\kappa =2 \cdot
10^6$m$^{-1}$, $\Sigma/\Sigma_0$ and
$\tilde{\sigma}/\tilde{\sigma}_0$ both become negative when the
membrane is conductive. No such change of sign is present in the
tension in the non-conductive {\it i.e.} capacitor limit when
$G=0$ (thick solid line). For clarity the horizontal solid line
represents the point of zero tension or zero of $\tilde{\sigma}$.
 (b) The ratio of normalized electrostatic
contribution to the bending modulus $K/K_0$ is shown as a function
of $\kappa$, where similarly $K$ is normalized by its value in the
infinite $\kappa$ limit $K_0$. The $G^*=0$ limit is represented as
a thick solid line and the $G^*=10 \Omega^{-1}$/m$^2$ case as a
dashed line.}
 \label{fig:tension-kappa}
\end{figure}

This point is illustrated in Figure~\ref{fig:tension-kappa}, where the
electrostatic contribution to the tension $\Sigma$ and the bending
modulus $K$, as a function of the inverse Debye length $\kappa$ in
the $G=0$ limit (solid line) and for $G^*=10 \Omega^{-1}$/m$^2$
(dashed line), are shown. The solid and dashed lines only deviate at small
values of $\kappa$. The decrease of $K/K_0$ with salt concentration
is also obtained in Ref.~\cite{leonetti}. We have no simple
explanation for the non-monotonicity of $K/K_0$ which is observed
near $\kappa=2 \cdot 10^6$m$^{-1}$, but note that a qualitatively
similar non-monotonicity - in the spontaneous curvature modulus,
however - is seen in Ref.~\cite{leonetti}.

We find a reversal of the sign of $\Sigma$ in the conductive case
at small values of $\kappa$. The sign reversal is absent in the
non-conductive case. This remarkable feature is shown in
Figure~\ref{fig:tension-kappa}(a). This mechanism of sign reversal
may provide an explanation of some recent experiments, such as the
study of cell movement of Ref.~\cite{sachs}, where a reversal of
movement/tension was observed in response to a change of ionic
strength. This change of sign of the tension is clearly due to a
change of sign of $\tilde{\sigma}$ as shown in Figure
\ref{fig:tension} (see also appendix for the condition of the
change of sign of $\tilde{\sigma}$). One can see there that a
change of sign of $\tilde{\sigma}$ occurs when the conductance $G$
or the parameter $\delta_m$ are varied away from the point where
$\tilde{\sigma}=0$, which occurs when $\delta_m \simeq 1/G$.

\begin{figure} {\par {
\rotatebox{0}{\includegraphics[scale=0.7]{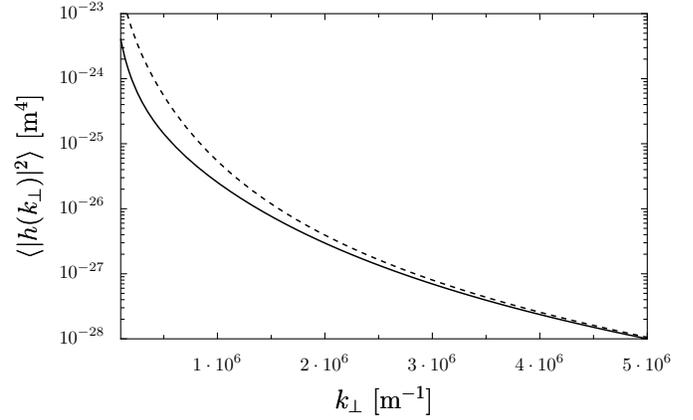}} }
\par}
\caption{Fluctuation spectrum of membrane fluctuations for the
numerical values of the parameters discussed in the text. The
solid line corresponds to the spectrum of a membrane of bare
tension $\sigma_0^*=10^{-7} $Jm$^{-2}$ and bare bending modulus
$\kappa_0^*=15 k_B T$, while the dashed line corresponds to the
fluctuation spectrum of a driven membrane in an electric field,
which we describe with equation~(\ref{fluctuation spectrum}). The
parameters are the same as discussed in the text except that here
the potential drop which is applied is only of $V^*=5$mV, so that
$\Sigma^*=-8.5 \cdot 10^{-8}$ Jm$^{-2}$, $\Gamma^* = 8.1 \cdot
10^{-22}$Jm$^{-1}$ and $K^* = 0.0011 k_B T$.}
 \label{fig:spectrum}
\end{figure}

\begin{figure} {\par {
\rotatebox{0}{\includegraphics[scale=0.7]{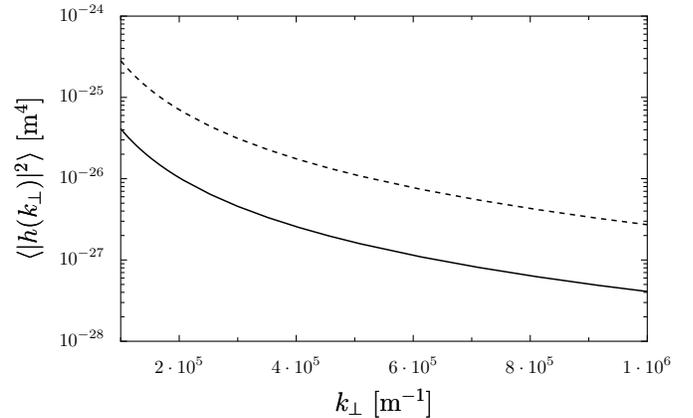}} }
\par}
\caption{Fluctuation spectrum of membrane fluctuations in the same
conditions as in Figure~\ref{fig:spectrum} except for the potential
$V^*=50$mV and for the bare tension $\sigma_0^*=10^{-5}
$Jm$^{-2}$.}
 \label{fig:spectrum2}
\end{figure}

Figures~\ref{fig:spectrum}-\ref{fig:spectrum2},
show fluctuation spectra corresponding to the parameter values as
indicated in the figure caption. The range of wavevectors
indicated corresponds roughly to experimentally accessible values
in video microscopy. Note the substantial increase of fluctuation
amplitudes at low wavevectors in Figure~\ref{fig:spectrum}, which
arises from the lowering of the surface tension. Such a lowering
is similar to the one observed in Ref.~\cite{PRL in preparation}.
For the parameters used in Figure~\ref{fig:spectrum2}, corresponding
to a larger bare tension $\sigma_0^*$, we see a significant
increase of fluctuations over the full range of wavevectors.

\section{Conclusion}
\label{conclusion} In conclusion, we have analyzed the steady state
fluctuations of a membrane driven by an applied DC electric field. Our
analysis is valid in the linear regime for the response of the ion
channels and for the description of the electrostatic effects. We have
confirmed the main results of Ref.~\cite{lacoste}, including the
presence of a term proportional to $\kps^3$ in the fluctuation
spectrum. We have provided a simple physical argument for the physics
underlying this term, relating it to a nonlinear electrokinetic effect
termed induced-charge electro-osmosis (ICEO). The predicted flow
around a curved driven membrane is in the reverse direction from
typical ICEO flows around blocking metal surfaces~\cite{iceo} and has
different dependence on the geometry, in the limit we describe for
the zero-thickness case. We stress, more generally, the
importance of electrokinetic effects such as the one described here
for descriptions of the dynamic properties of the soft,
non-equilibrium membranes found in living cells.

Although our calculations addressed the case of a membrane
driven out of equilibrium through the combined action
of an external potential and ion channels or pumps which
transferred ions across the surface, most of our results
should translate directly to the cell, where the potential
difference across the membrane is maintained solely through
active pumps and channels. This is a consequence of the
observation that once steady currents are established
which drive the system out of equilibrium, the precise
way in which such transmembrane currents are maintained
should be irrelevant to the description of
membrane properties, which are dominated by effects at
the much smaller scales of the bilayer thickness and the
Debye screening length.

We have also confirmed the importance of capacitive effects, which
are responsible for a negative contribution to the membrane
tension and can lead to membrane instabilities. In agreement with
the recent results of Ref.~\cite{lomholt}, we find that the
electrostatic and electrokinetic contribution to the bending
modulus increases with the salt concentration, whereas the
electrostatic contribution to the surface tension decreases with
the salt concentration. We have also found a reversal of sign of
the tension and of the surface charge in the Debye layer
$\tilde{\sigma}$ in the low salt limit (as compared to the
situation at high salt).

We have extended the calculations of Ref.~\cite{lacoste} by
including a channel concentration field in the description.  This
did not lead to qualitatively new effects within our perturbative
treatment with our assumptions that the response of the channels
to positive and negative ions is identical.

Extensions of the work described here include the
description of non-equilibrium effects in membranes
bearing a fixed charge which could be distributed
asymmetrically across the two layers. The modulation of
this fixed charge through remodeling of plasma membrane
lipids is now understood to play an important role in
cell division and phagocytosis.
The cytoskeleton of the cell couples
to the membrane, lending the coupled cytoskeleton-membrane
system a shear modulus. In addition,
cytoskeletal proteins are typically charged. Understanding how such
effects modify the elastic properties of the coupled
membrane-cytoskeleton system out of equilibrium is an
area which is largely unexplored.

The clarification of
the mechanical properties and fluctuation spectrum of
lipid vesicles containing active pumps and channels is
another possible application of the ideas presented here.
Incorporating a biologically more reasonable model for
the non-linear current voltage relation associated with
ion pumps into the calculation would provide a useful
extension of this work. Finally, achieving a more detailed
understanding of the role of non-linear electrokinetic
effects, such as ICEO, in modulating the dynamic properties
of membranes driven out of equilibrium, appears to be an
important new direction for further research.

\begin{acknowledgement}
We thank Patricia Bassereau, Madan Rao, Sriram Ramaswamy, Jacques
Prost, V. Kumaran, Pierre Sens and Armand Ajdari for useful
discussions. MZB thanks the National Science Foundation, under
Contract DMS-070764 for support. MZB also acknowledges the
hospitality of ESPCI and support from the Paris Sciences Chair. DL
and GIM acknowledge support from the Indo-French Center for the
Promotion of Advanced Research under Grant No. 3502, the DBT (India)
and the DST (India).
\end{acknowledgement}

\appendix
\section{Mapping the finite membrane thickness model to zero
thickness} The calculation of the electrostatic corrections in the
case of a membrane of finite thickness is complex, leading to
expressions which are often hard to interpret.  It is thus useful to
consider simpler, alternative formulations of the physics which could
be used to gain physical insight. We describe one such approach
briefly below, based on the ``Stern boundary condition'' for thin
dielectric layers~\cite{bazant2005}, and use it to calculate the
internal and external contributions to the surface tension.

The idea is to map the finite thickness problem into an equivalent
zero-thickness one, but with boundary conditions which are different
from the ones we considered in the body of the paper. Beginning
with the definition
\begin{equation} Q= \frac{1}{2} \left( \delta N_1-\delta
N_2 \right),
\end{equation}
following equations~(\ref{base state2})
we obtain
\begin{equation}
\partial_z^2 \Psi = -Q.
\end{equation}
We can write the solutions to the problem for the
$z >0$ case as
\begin{equation}
\Psi(z) = \sigma(z - \frac{L}{2}) - Ae^{-z} + \frac{V}{2},
\end{equation}
and for the $z < 0$ case
\begin{equation}
\Psi(z) = \sigma(z + \frac{L}{2}) - A^{'}e^z - \frac{V}{2}.
\end{equation}
Applying the electrostatic boundary condition
\begin{equation}
r \partial_z \Psi_m(\pm \frac{d}{2}) =
\partial_z \Psi(z = \pm\frac{d}{2})
\end{equation}
where the internal electric field is
\begin{equation}
E_m = -\partial_z \Psi_m(z).
\end{equation}
The internal field is a constant which we can calculate from
\begin{equation}
\int_{-d/2}^{d/2} E_m (z) dz = d E_m = \Psi(-\frac{d}{2}) -
\Psi(\frac{d}{2}).
\end{equation}
This leads to the boundary condition
\begin{equation}
- \partial_z \Psi(z = \pm\frac{d}{2}) = \frac{r}{d}\left [
\Psi(-\frac{d}{2}) - \Psi(\frac{d}{2}) \right ].
\end{equation}
We will now treat this as an equivalent zero thickness problem
with the constraint that
\begin{equation}
\delta_m \partial_z \Psi(z = 0^\pm)
= \Psi(0^+) - \Psi(0^-)   \label{eq:stern}
\end{equation}
where
\begin{equation}
\delta_m =\frac{d}{r}=\frac{\epsilon d}{\epsilon_m}
\end{equation}
is an effective length scale characterizing the membrane (scaled
to $\kappa^{-1}$), over which the potential in the electrolyte
extrapolates linearly to its value on the other side of the
membrane. Note that equation~(\ref{eq:stern}) is a mixed Robin-type
boundary condition, involving both the field $\Psi$ and its
derivative at the boundaries $z = 0^\pm$. In order to pass to the
limit of zero membrane thickness, $d = d^* \kappa \rightarrow 0$,
we take the joint limit $r = \epsilon_m/\epsilon \rightarrow 0$,
keeping $\delta_m$ fixed. The boundary condition (\ref{eq:stern})
also explicitly shows the importance of the coupling parameter $t
= \delta_m^{-1}$, which has been discussed for finite-thickness
membranes~\cite{helfrich,andelman,kleinert,chou}.

The boundary condition equation (\ref{eq:stern}) is now widely used to
describe thin dielectric layers on metal surfaces and
electrodes~\cite{bazant2004,bazant2005}, although we are not aware
of any prior application to ion-permeable membranes.  It was
perhaps first used to describe the compact Stern layer at the
electrode/electrolyte interface~\cite{itskovich} and recently
extended to nonlinear surface capacitance~\cite{bazant2005}. In
this context, it has been postulated that the field-dependent
voltage drop, $\Delta\psi=\delta\, \partial_z \psi$, drives
Faradaic electrochemical reactions~\cite{bonnefont,bazant2005}. In
our case, the same voltage drop contributes to electrochemical
potential differences across the membrane, which set the ionic
currents. In modeling ICEO flows around metal surfaces, the same
boundary condition is also used to describe thin dielectric
coatings, such as oxide layers~\cite{ajdari,iceo,levitan}, which
is again similar to our modeling of ICEO flow around a driven
membrane. The same type of Robin-type boundary condition has also been derived under more general conditions for the interface between a dielectric body (not necessarily a thin layer) and an electrolyte and used to model ICEO flows around dielectric microchannel corners and dielectric particles~\cite{yossifon}.

The physical interpretation of the parameter $\delta_m$ becomes more
clear when written in terms of dimensional variables,
\begin{equation}
\delta_m = \frac{\epsilon \kappa}{\epsilon_m/ d^*} =
\frac{C_D}{C_m}
\end{equation}
as the ratio of the low-voltage capacitance of the diffuse part of
the double layer, $C_D = \epsilon \kappa$, to that of the compact
part, $C_m = \epsilon_m/d^*$, which in our case is the membrane
(but could also be a surface coating or Stern layer). In the
linear regime of low voltages ($< kT/e$) and for thin double
layers ($\kappa L \gg 1$, $d=d^* \kappa \gg 1$), these are
constant capacitances, effectively in series~\cite{bazant2004},
where $\delta_m(1+\delta_m)^{-1}$ is the fraction of the total
double-layer voltage across the membrane, while
$(1+\delta_m)^{-1}$ is the fraction across the diffuse layers.

There are two limiting cases of (\ref{eq:stern}) which are
commonly assumed in the literature~\cite{bazant2005}. In the
``Gouy-Chapman limit'' $\delta_m \ll 1$, most of the voltage drop
occurs in the diffuse layer. In the ``Helmholtz limit'' $\delta_m
\gg 1$, the compact layer -- or in our case, the membrane --
carries most of the voltage. We make the latter assumption in the
main text to reduce equation (\ref{eq:stern}) to the simpler boundary
condition of equation~(\ref{BC3b}).

Here, we briefly consider the general case $0<\delta_m <
\infty$.  Using the above equations,
\begin{eqnarray}
\partial_z \Psi(z = 0^+) = \sigma + A \nonumber, \\
\partial_z \Psi(z = 0^-) = \sigma - A^{'},
\label{Electric field BC}
\end{eqnarray}
we have thus
\begin{equation}
\partial_z \Psi(z = 0^+) = \partial_z \Psi(z = 0^-) \implies A = -A^{'}
\end{equation}
We can fix $A$, using the result for $\Psi(z)$,
yielding
\begin{equation}
A = \frac{V + \sigma(-\delta_m - L)}{2 + \delta_m}. \label{A}
\end{equation}
Comparing equations~(\ref{Electric field BC}) with
equation~(\ref{Efield-finite d}), we see that $A=\tilde{\sigma}$
and equation~(\ref{A}) are equivalent to equation~(\ref{rel
sigmas}) in the limit $r \rightarrow 0, d \rightarrow 0$, keeping
$\delta_m$ fixed.

We now illustrate the calculation of the tension, using our earlier result
\begin{eqnarray}
\Sigma^{out} &=&-\int_{-L/2}^{L/2}
(E_z^{(0)})^2(z) dz \nonumber \\
&+& \frac{L}{2} \left[
  (E_z^{(0)})^2(z \rightarrow \infty) + (E_z^{(0)})^2(z \rightarrow -\infty)
  \right].
\end{eqnarray}
which yields
\begin{eqnarray}
\Sigma^{out} &=&-2\int_{0}^{L/2} E^2(z) dz +L E^2(z \rightarrow \infty) \nonumber \\
&=& -2\int_{0}^{L/2} (\sigma + Ae^{-z})^2 dz + L \sigma^2 \nonumber \\
&=& -2\int_{0}^{L/2} (\sigma^2 + A^2e^{-2z} +2 \sigma A e^{-z}) dz + L \sigma^2 \nonumber \\
&=& -2 \left [ \frac{A^2}{2} + 2\sigma A \right ],
\end{eqnarray}
which goes to $3 \sigma^2$ in the limit of $\delta_m \rightarrow 0$
for the zero thickness limit, where $A=-\sigma=\tilde{\sigma}$. We can
also obtain \bb \Sigma^{in} = -r E_m^2 = \frac{-r}{d^2}\left[
  \Psi(0^+) -\Psi(0^-)\right ]^2 = - \frac{d}{r} (\sigma +A)^2 \en
which coincides with equation~(\ref{sigma_in}) given in the main
text.

Using the definition of $\sigma$ of equation (\ref{def_sigma}),
and the expression of the ion fluxes of Eqs.~(\ref{final
currents})-(\ref{final currents2}), we have that $\sigma \simeq
GV/(1+GL)$. From equation (\ref{A}), one obtains \bb A \simeq
\frac{V}{\left( 1+GL \right) \left( 2 + \delta_m \right)}
\left(1-G \delta_m \right), \en which shows that $A>0$ when
$\delta_m < 1/G$ and $A<0$ when $\delta_m > 1/G$. The condition
$\delta_m < 1/G$ is equivalent to $\epsilon_m/\epsilon \gg
G^*/G_0^*$, where $G^*$ is the typical conductance of typical ion
channels/pumps defined in equation (\ref{conductance}) and
$G_0^*=Dn^*e^2/d^* k_B T$ is the conductance of a layer of
electrolyte of thickness $d^*$.

\section{Solution of the Stokes equations for the first model of a membrane
of zero thickness and zero dielectric constant}
We recall that the normal component of the velocity satisfies a single fourth order differential equation, Eq.~\ref{4order}. With the expressions for the charge and the potential at zeroth and first order given in the previous sections, the flow can be solved on each side separately as
\begin{align*}
& v_z(\kp,z)  =  \left( A_1 + B_1 z \right) e^{-\kps z} + C_1 e^{-l z}   & \mbox{for } & 0<z   \ ,\\
& v_z(\kp,z)  =  \left( A_2 + B_2 z \right) e^{\kps z} + C_2 e^{l z}   & \mbox{for } & z<0 \ ,
\end{align*}
where the integration constants must be determined by proper matching boundary conditions.
Note that the boundary conditions on the membrane are enforced at $z=0^\pm$ rather than at the actual position $h(\vr)$ of the interface because of our assumption
of small deformations limited to first order in the membrane height.

Imposing the boundary conditions for the velocity of Eqs.~\ref{permeation}-\ref{continuity dvz/dz}, the flow on the positive side $z>0$ is explicitly calculated to be
\begin{eqnarray} \label{explicitvz}
v_z(\kp,z) & = & \sigma^2 \kps^2 \left( z-\frac{1}{l} - \frac{z \kps}{l} \right) e^{-\kps z} h(\kp) \\
           & + & s \left( 1 + z \kps \right) e^{-\kps z} h(\kp) \\
           &+ & \frac{\sigma^2 \kps^2}{l} e^{-l z} h(\kp),
\end{eqnarray}
and
\begin{eqnarray} \label{explicitvx}
\vp (\kp,z) & = & -i \kp  \sigma^2 \left( -1+ \kps z - \frac{\kps^2 z}{l} \right) \\
            & \times  &  e^{-\kps z} h(\kp) - zs  i \kp h(\kp)e^{-\kps z} \\
            & -  &   i \sigma^2 \kp h(\kp) e^{-lz}.
\end{eqnarray}
Note that, at this point, the stress boundary conditions have not been used yet.
In the particular case where no electrostatic force is present (for $\sigma=0$), one recovers the fluid flow created with a membrane bending mode \cite{levine}, which is represented in Fig.~\ref{fig:hydro flows}a.
\begin{eqnarray}
v_z(\kp,z) & = & s \left( 1+ z \kps \right) e^{-\kps z} h(\kp), \\
\vp(\kp,z) & = & -i \kp z s e^{-\kps z} h(\kp).
\end{eqnarray}

In the general case where $\sigma \neq 0$, we are interested in the solution of the Stokes equation
where the growth rate $s$ is determined from the stress boundary conditions. For Fig.~\ref{fig:hydro flows}b and Figs.~\ref{fig:thickDLflow}, we have assumed that $s = 0$, which corresponds to a quasi-stationary membrane, whose shape is determined by the flow field.

The stress component along $z$, obtained to first order in the
membrane height field and evaluated at the membrane surface, is
\bb \tau_{zz}^{(1)}=-p+2
\partial_z v_z + \partial_z \psi^{(1)} \partial_z \Psi + h(\rp)
\partial_z \, [ -P + (\partial_z \Psi)^2/2
]. \en As follows from equations~(\ref{E+},\ref{E-}), and (\ref{Pressure}), the
stress is balanced in the base state. Thus, the last term drops out
and \bb \tau_{zz}^{(1)}= \left( -p+2
\partial_z v_z + \partial_z \psi^{(1)} \partial_z \Psi \right)_{z=0}. \en
Similarly the transverse stress is \bb \tau_{\perp z}^{(1)}=
\left(
\partial_\perp v_z + \partial_\perp \psi^{(1)} \partial_z \Psi \right)_{z=0}. \en

In fact, because of our use of the boundary condition of a
vanishing electric field on the membrane, these expressions further simplify to
$ \tau_{zz}^{(1)}= \left( -p+2
\partial_z v_z \right)_{z=0}$ and $\tau_{\perp z}^{(1)}=
\left(
\partial_\perp v_z \right)_{z=0}.$
These stresses can be evaluated using the expression of the pressure in terms of $v_z$ and $\fp$, while $v_z$ itself can be obtained by solving Eq.~\ref{4order}.

With the expressions of the velocity given in Eqs.~\ref{explicitvz}-\ref{explicitvx}, the discontinuity in the normal-normal stress Eq.~\ref{BC stress normal} component fixes the value of the growth rate $s$. After expanding the obtained expression in powers of $\kps$, one obtains the growth rate equation given in Eq.~\ref{eq motion 2}.

After inserting the expression of the growth rate $s$ into the equations for the flow field given in Eqs.~\ref{explicitvz}-\ref{explicitvx} and Taylor expanding with respect to $\kps$,
one finds that (at lowest order in $\kps$)
\bb
\vp (\kp, z) \simeq i \kp h(\kp) \sigma^2 \left( 1 - e^{-z} \right),
\en
which is essentially the result of Eq.~\ref{approx vperp}. This result confirms that the tangential  velocity $\vp$, which is strictly zero at $z=0$ according to the non-slip boundary condition, has a significant (non-zero) value at a distance $z$ of the order of one Debye length away from the interface, as predicted from the Helmholz-Smoluchowski formula of
Eq.~\ref{Helmholz-Smoluchowski}.

\end{document}